# CLUSTERED FLEXIBLE CALIBRATION PLOTS FOR BINARY OUTCOMES USING RANDOM EFFECTS MODELING

Lasai Barreñada[1,2], Bavo D.C. Campo[1,3,4], Laure Wynants[1,2,5], Ben Van Calster[1,2]

[1] Department of Development and Regeneration, KU Leuven, Belgium

[2] Leuven Unit for Health Technology Assessment Research (LUHTAR), KU Leuven, Belgium

[3] Department of Metabolism, Digestion and Reproduction, Imperial College, United Kingdom.

[4] Department of Accountancy, Finance and Insurance, KU Leuven, Belgium

[5] Department of Epidemiology, CAPHRI Care and Public Health Research Institute, Maastricht University, Maastricht, Netherlands.

**Corresponding author**

Ben Van Calster

KU Leuven, Department of Development and Regeneration

Herestraat 49 box 805

3000 Leuven

Belgium

ben.vancalster@kuleuven.be

Evaluation of clinical prediction models across multiple clusters, whether centers or datasets, is becoming increasingly common. A comprehensive evaluation includes an assessment of the agreement between the estimated risks and the observed outcomes, also known as calibration. Calibration is of utmost importance for clinical decision making with prediction models and it may vary between clusters. We present three approaches to take clustering into account when evaluating calibration. (1) Clustered group calibration (CG-C), (2) two-stage meta-analysis calibration (2MA-C) and (3) mixed model calibration (MIX-C) can obtain flexible calibration plots with random effects modelling and providing confidence and prediction intervals. As a case example, we externally validate a model to estimate the risk that an ovarian tumor is malignant in multiple centers (N = 2489). We also conduct a simulation study and synthetic data study generated from a true clustered dataset to evaluate the methods. In the simulation study MIX-C and 2MA-C (splines) gave estimated curves closest to the true overall curve. In the synthetic data study MIX-C produced cluster specific curves closest to the truth. Coverage of the prediction interval across the plot was best for 2MA-C with splines. We recommend using 2MA-C with splines to estimate the overall curve and the 95% PI and MIX-C for the cluster specific curves, especially when sample size per cluster is limited. We provide ready-to-use code to construct summary flexible calibration curves with confidence and prediction intervals to assess heterogeneity in calibration across datasets or centers.

# Highlights

**What is already known**

**Traditional methods for assessing calibration in clinical prediction models often assume independence between observations. However, when performing an external validation where the data are clustered, this assumption is violated. Ignoring clustering can impair the reliability of calibration assessments, potentially leading to misguided clinical decisions.**

**What is new**

**We introduce three novel methodologies that explicitly account for clustering in calibration assessments during clinical prediction model validation. These methods generate an overall calibration curve with prediction intervals, representing expected calibration curves in hypothetical new clusters. Among these, the mixed model calibration (MIX-C) method is particularly recommended, as it provides cluster-specific calibration curves, and the two-stage meta-analysis calibration (2MA-C) with splines, as it provides good overall calibration and prediction intervals with decent coverage. To facilitate adoption, we provide ready-to-use R functions.**

**Potential impact for Research Synthesis Methods readers**

**External validations aim to assess how clinical prediction models perform across diverse external settings. Neglecting the clustered nature of data in multicenter validations or meta-analyses of validation studies undermines calibration analysis, overlooking cluster-specific insights. The methodologies we propose address these limitations by estimating overall calibration curves and offering prediction intervals for hypothetical new clusters. This approach enables more refined and reliable calibration assessments, directly supporting informed decision-making in clinical research.**

# 1 INTRODUCTION

Clinical prediction models (CPMs) are evidence-based tools that estimate the probability of health-related events either at the time of evaluation (diagnosis) or at some point in the future (prognosis).[1,2] These risk estimates play a crucial role in evidence-based decision-making and can be of great value when counselling patients. However, to ensure the reliability of risk estimates, CPMs must exhibit good calibration.[3] Calibration assesses how well the predicted risks correspond to observed risks.[4] In recent years, there has been a notable increase in studies involving multiple clusters. The term "cluster" can refer to different groupings within the population under study, depending on the context. For example, in multicenter studies, a cluster commonly refers to different centers, whereas in meta-analysis, a cluster may pertain to each study or centers within the study where possible (e.g. individual patient data (IPD) meta-analysis[5]). In this work, the term "cluster" will specifically denote a center in a multicenter study, but the methods proposed apply more generally as well. An analysis of CPMs included in Tufts PACE (Predictive Analytics + Comparative Effectiveness), developed after the year 2000, revealed that 64% of the models utilized multicenter (clustered) data, indicating a growing trend in such studies.[6,7] TRIPOD (Transparent Reporting of a Multivariable Prediction Model for Individual Prognosis or Diagnosis) has addressed this development by publishing an extension known as TRIPOD-Cluster.[8] This extension underscores the importance of taking clustering into account when developing or evaluating prediction models.

Miscalibration can have a harmful influence on medical decision making.[3] It is common to observe heterogeneity in model performance across centers or studies, most notably in model calibration.[9–12] It can thus be misleading to generalize performance results from single-cluster data.[13–16] For these reasons, it is crucial to collect clustered data to investigate and quantify heterogeneity in calibration performance between clusters. The presence of clustering introduces challenges, as individuals are no longer independent due to correlations among patients within the same cluster.[17,18] Traditional (non-clustered) methodologies should therefore be adapted for use with clustered data.

The most informative assessment of calibration performance for a prediction model is a calibration plot, used to assess whether, among patients with the same estimated risk of the event, the observed proportion of events equals the estimated risk.[4,19]A calibration plot has estimated risks on the x-axis, and the observed proportion on the y-axis. For binary outcomes, observed outcome values are $Y = 0$ (no event) or $Y = 1$ (event). The observed proportion of individuals with the event conditional on estimated risk are not observed directly, but are estimated. One approach consists of creating $Q$ groups (usually based on quantiles of estimated risks) and estimating the observed proportion in each of the groups as the proportion of individuals with the condition. We refer to this approach throughout the text as grouped calibration. The calibration plot then has $Q$ dots representing the mean risk (value on the x-axis) and observed proportion (value on the y-axis). Another approach is using a flexible model.[20] The observed outcome is regressed on the estimated probabilities using a smoother such as local regression (loess) or splines. The smoothed relation or $Q$ groups are plotted together with a line of identity which represents perfect calibration. The grouped approach loses information by categorizing the estimated risks, and depends on the value of $Q$ and the grouping approach. Flexible models are dependent on smoothing parameters. Flexible calibration curves have been presented for different problems like survival models[21], competing risk models[22], or multiclass models.[23] Flexible calibration curves that consider clustering have been understudied. Although summarizing statistics of calibration (e.g. O:E) can be meta-analysed across clusters, such summarizing statistics are by definition less informative than the calibration plot.[19,24] We present three different approaches to construct flexible calibration curves from clustered data, provide cluster-specific curves, and quantify heterogeneity using prediction intervals. We illustrate the methodologies with a real case example on ovarian cancer data and compare the performance of the methodology using simulated data, and synthetic data. Section 2 presents the motivating example, the general notation used through the paper, and introduces the three approaches to obtain calibration plots accounting for clustering. Section 3 and 4 present the methodology and results of the simulation study and synthetic data analysis, respectively, and in section 5 we conclude by discussing our findings. This study is an initial assessment of methods for clustered calibration analysis that we classify as a phase 2 methodological study in the four phase framework from Heinze and colleagues.[25]

# 2 METHODS

## 2.1 MOTIVATING EXAMPLE: ADNEX MODEL AND OVARIAN CANCER DATA

The methods will be illustrated using the ADNEX model for ovarian tumor discrimination that was developed with data from the International Ovarian Tumor Analysis (IOTA) group.[26,27] The ADNEX model is a multinomial regression mixed model with random intercepts per center to estimate the probability that an ovarian mass is malignant. Predicting malignancy of ovarian masses prior to surgery is important because benign masses can be managed conservatively, and malignant masses require different surgical approaches depending on the malignant subtype. ADNEX estimates the risk of five outcomes: benign tumor, borderline tumor, stage I invasive cancer, stage II-IV invasive cancer, and secondary metastasis. This work will focus on the overall risk of malignancy, which is obtained by adding the risks for the malignant subtypes. ADNEX has three clinical and six ultrasound predictors: age, serum CA125 level, type of center (oncology center or other hospital type), maximal diameter of the lesion, proportion of solid tissue, number of papillary projections, presence of >10 locules, acoustic shadows, and ascites. CA125 is optional, and in this work we focus on ADNEX without CA125. ADNEX coefficients already capture some of the between-center heterogeneity through the type of center. To illustrate multicenter calibration of the ADNEX model we use an external validation dataset of 2489 patients recruited in 17 hospitals.[28] Previous external validation of ADNEX on this dataset suggested that there was important heterogeneity between hospitals (**Figure S1**).[9] To avoid computation errors, three small non-oncology centers in Italy were combined, as well as two small non-oncology centers in the United Kingdom. The dataset then contains 14 clusters, with a median number of 189 patients per cluster (range 38 to 360) (**Figure 1**). The mean estimated probabilities (range 0.10 to 0.53) and the prevalence of malignancy (range 0.16 to 0.72) varied between clusters. The intra-cluster correlation (ICC) in (logit) malignancy risk based on the null random intercept model is 15%.[17] An ICC of 0 would mean that all clusters are similar, and all variance in the logit malignancy risk is due to individual-level variability.

## 2.2 NOTATION

We assume to have a dataset with a total of $J$ clusters and use $j = (1, \dots, J)$ to index the clusters. In each cluster, we have $n_j$ patients and we index the patients using $i = (1, \dots, n_j)$. The total sample size is $N = \sum_{j=1}^{J} n_j$. We use $y_{ij} \in 0,1$ to denote the outcome of patient $i$ in cluster $j$, which takes on the value 0 in case of a non-event and 1 in case of an event. We assume that $y_{ij}$ follows a Bernoulli distribution $y_{ij} \sim \text{Bern}(\pi_{ij})$, where $\pi_{ij}$ denotes the probability of experiencing the event. Typically, $\pi_{ij}$ is expressed as a function of a set of risk characteristics, captured in the covariate vector $\boldsymbol{x}_{ij}$ and we assume that there exists an unknown regression function $r(\boldsymbol{x}_{ij}) = P(y_{ij} = 1 | \boldsymbol{x}_{ij})$. We approximate this function using a risk prediction model, where we model the outcome as a function of the observed risk characteristics and employ statistical or machine learning techniques to estimate this model. A general expression that encompasses both types is

$$\mathrm{P}(\, y_{ij} = 1 \mid \boldsymbol{x}_{ij} \,) = \pi(\boldsymbol{x}_{ij}) = f(\boldsymbol{x}_{ij}) \quad (1)$$

Where $f(\cdot)$ denotes the predicted risk given covariate vector $\boldsymbol{x}_{ij}$. In a logistic regression model, we have that

$$\pi(\boldsymbol{x}_{ij}) = \frac{e^{\boldsymbol{x}_{ij}^T\boldsymbol{\beta}}}{1 + e^{\boldsymbol{x}_{ij}^T\boldsymbol{\beta}}}$$

where $\boldsymbol{\beta}$ denotes the parameter vector. We can rewrite the above equation to

$$log\left(\frac{\pi(\boldsymbol{x}_{ij})}{1 - \pi(\boldsymbol{x}_{ij})}\right) = \boldsymbol{x}_{ij}^T\boldsymbol{\beta}$$
$$\text{logit}\left(\pi(\boldsymbol{x}_{ij})\right) = \boldsymbol{x}_{ij}^T\boldsymbol{\beta}$$

We can employ splines to allow for a non-linear relationship between the covariates and outcome.[29] To estimate equation (1), we fit a statistical or machine learning model to the training data. The resulting model fit then provides us with the predicted probability $\hat{\pi}(\boldsymbol{x}_{ij})$.

Using calibration curves, we examine how well the predicted probabilities correspond to the actual event probabilities.[1,4,30,31] A calibration plot maps the predicted probabilities $\hat{\pi}(\boldsymbol{x}_{ij})$ to the actual event probabilities $\text{P}\left(y_{ij} = 1 \,\middle|\, \hat{\pi}(\boldsymbol{x}_{ij})\right)$ and hereby provides a visual representation of the alignment between the model's estimated risks and the true probabilities. For a perfectly calibrated model, the calibration curve follows the diagonal as $\text{P}\left(y_{ij} = 1 \,\middle|\, \hat{\pi}(\boldsymbol{x}_{ij})\right) = \pi(\boldsymbol{x}_{ij})$ for all $i$.

## 2.3 FLEXIBLE CALIBRATION PLOTS IGNORING CLUSTERING

We can estimate the calibration curve using a logistic regression model [31–33]

$$\text{logit}\left(\text{P}\left(y_{ij} = 1 \,\middle|\, \hat{\pi}(\boldsymbol{x}_{ij})\right)\right) = \alpha + \zeta\, \text{logit}\left(\hat{\pi}(\boldsymbol{x}_{ij})\right)$$

where we estimate the observed proportions as a function of the (logit transformed) predicted probabilities. Using the fitted model, we create a calibration curve by plotting the $\hat{\pi}(\boldsymbol{x}_{ij})$ against the observed proportions $\hat{P}\left(y_{ij} = 1 \,\middle|\, \hat{\pi}(\boldsymbol{x}_{ij})\right)$. This model, however, only allows for a linear relationship and, as such, does not adequately capture moderate calibration. To allow for a non-linear relationship between $\hat{\pi}(\boldsymbol{x}_{ij})$ and $y_{ij}$, we can rely on non-parametric smoothers, such as locally estimated scatterplot smoothing (loess) or restricted cubic splines

$$\text{logit}\left(\text{P}\left(y_{ij} = 1 \,\middle|\, \hat{\pi}(\boldsymbol{x}_{ij})\right)\right) = s\left(\text{logit}\left(\hat{\pi}(\boldsymbol{x}_{ij})\right)\right).$$

Here, $s(\cdot)$ denotes the smooth function applied to the logit transformed predicted probability. This results in a flexible calibration plot, which is implemented in R packages such as `CalibrationCurves` (`val.prob.ci.2` function), `rms` (`val.prob` function), or `tidyverse` (`cal_plot_logistic` function).[4(p20),29,34,35]

**Figure 2** presents a flexible calibration curve (using a restricted cubic spline) based on the motivating example with pooled data, and the results of a grouped calibration assessment for ovarian malignancy prediction. The panel below shows the sample distribution of the estimated risks for benign cases (blue) and malignant cases (red). The plots suggest that risks were estimated too low.

Since there is no commonly accepted way to estimate calibration curves for prediction models in clustered data, we present three approaches covering different statistical methodologies. First, we present a grouped clustered calibration plot based on a bivariate random effects meta-analysis model (Clustered Group Calibration; CG-C). Second, we introduce a two-stage univariate random effects meta-analytical approach for the estimation of the observed events (Two-stage meta-analysis calibration; 2MA-C). Finally, we provide a one-step approach where we fit a random effects model with smooth effects to obtain individual calibration slopes per cluster (Mixed model calibration; MIX-C). We present the summary of the three proposed approaches in **Table 1.**

## 2.4 CLUSTERED GROUP CALIBRATION (CG-C)

The CG-C approach is a two-stage approach that extends the traditional grouped calibration approach to clustered data. In the traditional approach, data is pooled and groups are created based on quantiles (often deciles). In CG-C, we create $Q$ quantiles in every cluster (based on the estimated risk distribution per cluster) with $q = (1, \ldots, Q)$. Hereafter, we pool the estimated risks and observed proportions with a bivariate random effects meta-analysis at each quantile $q$. The logit-transformed observed proportion and average estimated risk are the response variables and we capture the cluster-specific deviation by including a random intercept. The equation of this model is given by

$$\begin{bmatrix} \text{logit}(\bar{y}_{qj}) \\ \text{logit}(\bar{\bar{\pi}}_{qj}) \end{bmatrix} = \begin{bmatrix} {}_{\bar{y}}\mu_q + u_{qj} + \varepsilon_{qj} \\ {}_{\bar{\bar{\pi}}}\mu_q + \nu_{qj} + \epsilon_{qj} \end{bmatrix} \tag{2}$$

where $\bar{y}_{qj}$ denotes the prevalence of cluster $j$ within quantile $q$ and $\bar{\bar{\pi}}_{qj}$ the average estimated risk in cluster $j$ within quantile $q$. We use the subscript $q$ in the quantities to indicate that this refers to quantile $q$. $u_{qj}$ and $\nu_{qj}$ represent the random intercepts for cluster $j$, capturing the between-cluster heterogeneity whereas $\varepsilon_{qj}$ and $\epsilon_{qj}$ account for the within-cluster error. In this model, we assume that ${}_{\bar{y}}\mu_q + u_{qj}$ and ${}_{\bar{\bar{\pi}}}\mu_q + \nu_{qj}$ are randomly drawn from a distribution with mean (or pooled) ${}_{\bar{y}}\mu_q$ and ${}_{\bar{\bar{\pi}}}\mu_q$, respectively. Additionally, we assume that

$$\begin{bmatrix} u_{qj} \\ \nu_{qj} \end{bmatrix} \sim \mathcal{N}\left(\begin{bmatrix} 0 \\ 0 \end{bmatrix}, \Omega_q\right) \text{ and } \begin{bmatrix} \varepsilon_{qj} \\ \epsilon_{qj} \end{bmatrix} \sim \mathcal{N}\left(\begin{bmatrix} 0 \\ 0 \end{bmatrix}, \Sigma_{qj}\right)$$

where $\Omega_q$ and $\Sigma_{qj}$ represent the between-cluster and within-cluster covariance matrices within quantile $q$, respectively.

$$\Omega_q = \begin{bmatrix} {}_{\bar{y}}\tau_q^2 & {}_{b}\rho_q \, {}_{\bar{y}}\tau_q \, {}_{\bar{\bar{\pi}}}\tau_q \\ {}_{b}\rho_q \, {}_{\bar{y}}\tau_q \, {}_{\bar{\bar{\pi}}}\tau_q & {}_{\bar{\bar{\pi}}}\tau_q^2 \end{bmatrix}$$

$$\Sigma_{qj} = \begin{bmatrix} {}_{\bar{y}}\sigma_{qj}^2 & {}_{w}\rho_q \, {}_{\bar{y}}\sigma_{qj} \, {}_{\bar{\bar{\pi}}}\sigma_{qj} \\ {}_{w}\rho_q \, {}_{\bar{y}}\sigma_{qj} \, {}_{\bar{\bar{\pi}}}\sigma_{qj} & {}_{\bar{\bar{\pi}}}\sigma_{qj}^2 \end{bmatrix}$$

We use ${}_{\bar{y}}\sigma_{qj}^2$ and ${}_{\bar{\bar{\pi}}}\sigma_{qj}^2$ to denote the variance of $\bar{y}_{qj}$ and $\bar{\bar{\pi}}_{qj}$, ${}_{w}\rho_q$ represents the within-cluster correlation and ${}_{b}\rho_q$ the between-cluster correlation in quantile $q$. The between-study variance is denoted as ${}_{\bar{y}}\tau_q^2$ and ${}_{\bar{\bar{\pi}}}\tau_q^2$. For each quantile *q,* we use a bivariate meta-analysis model to account for the strong dependence between the average estimated risk and observed proportion. The calibration plot is then created by plotting the $Q$ values of ${}_{\bar{\bar{\pi}}}\mu_q$ in the x-axis and the ${}_{\bar{y}}\mu_q$ in the y-axis. This approach has the advantage of obtaining an observed proportion per cluster in a model-agnostic way. In addition, it provides uncertainty measures in the cluster with the average random effect with confidence intervals for that effect and in a hypothetical new cluster (PI; prediction intervals). It also has some limitations. First, it has the same drawback as the traditional grouped calibration because the curve is dependent on the number of quantiles selected, especially when the sample size is limited. Second, and linked to the previous limitation, the arbitrary selection of $Q$ quantiles might create groups that are too heterogeneous between clusters because the estimated risks distributions differ (e.g. a high risk setting vs a low risks setting). Third, the meta-analysis does not provide cluster-specific curves using empirical Bayes. We used the `rma.mv` function of the `metafor` package.[36]

We include an extension to the proposed methodology that we call "interval" grouping. The methodology is the same except for the way the groups are created. Per cluster, we create $I$ groups by dividing the probability space (0-1) into $I$ equally spaced intervals. In this way, we will create $I$ groups of different sizes based on the estimated risks. By doing this, we reduce the within group variability at the expense of not having necessarily all clusters present in all $I$ groups. The algorithm, additional details on the implementation and illustration are available in the **Supporting material (Appendix A1, Figure S2-S5) and code for the implementation in the OSF repository.**[37]

## 2.5 Two-stage meta-analysis calibration (2MA-C)

For the second approach, we use a two-stage random effects meta-analysis to estimate the calibration plot. First, we fit a flexible curve per cluster with a smoother of choice. Currently, we have implemented restricted cubic splines and LOESS. For LOESS, in each cluster, the span parameter with the lowest bias-corrected AIC is selected. For restricted cubic splines, the number of knots per cluster is selected by performing a likelihood ratio test between all combinations of models with 3,4, and 5 knots, selecting the model with the least knots that provides the best fit. We train a flexible curve per cluster and estimate the observed proportion for a fixed grid of estimated risks from 0.01 to 0.99 ($G$, default is 100 points). Hereafter, we use a random effects meta-analysis model per point in the grid to combine the logit-transformed predictions across clusters, fitting the following univariate model per point $g$ in the grid.

$$\text{logit}({}_s\widehat{\pi}_{gj}) = {}_{\widehat{\pi}}\mu_g + \nu_{gj} + \epsilon_{gj}.$$

${}_s\widehat{\pi}_{gj}$ denotes the predicted proportion of point $g$ for cluster $j$, ${}_{\widehat{\pi}}\mu_g$ the overall mean for point $g$, $\nu_{gj}$ the cluster-specific deviation and $\epsilon_{gj}$ the error. We assume that $\nu_{gj} \sim \mathcal{N}(0, {}_{\widehat{\pi}}\tau_g^2)$ and $\epsilon_{gj} \sim \mathcal{N}(0, {}_{\widehat{\pi}}\sigma_{gj}^2)$. Further, we include the inverse of the variance of $\text{logit}({}_s\widehat{\pi}_{gj})$ as weight and hereby take both the cluster-specific (${}_{\widehat{\pi}}\sigma_{gj}^2$) and between-cluster variability (${}_{\widehat{\pi}}\tau_g^2$) into account. As such, clusters with more precise estimates are assigned greater weights. This approach has the strength of being easy to compute and providing heterogeneity measures for the cluster with average calibration. This allows to plot confidence and prediction intervals which help to visually assess the certainty of the average curve (CI) and the heterogeneity between hypothetical new clusters (PI) in the whole range of predicted probabilities. The main limitation is that it is based on the smoothing technique used in each individual cluster so the curves will vary depending on the smoother selected. Additionally, the technique treats each point in the grid as independent, leading to pointwise confidence and prediction intervals. Cluster-specific curves are obtained in the first stage independently from the rest of the clusters. The algorithm, additional details on the implementation and illustration are available in the **Supporting material (Appendix A2, Figure S6-S7) and code for the implementation in the OSF repository.**

## 2.6 Mixed model calibration (MIX-C)

In the third approach, we employ a one-stage logistic generalized linear mixed model (GLMM) to model the outcome as a function of the logit-transformed predictions. To allow for a non-linear effect, we employ restricted cubic splines with three knots for both the fixed and random effects

$$\text{logit}\left(\text{P}\left(y_{ij} = 1 \mid \widehat{\pi}(\boldsymbol{x}_{ij}), \tilde{s}_j\right)\right) = s\left(\text{logit}\left(\widehat{\pi}(\boldsymbol{x}_{ij})\right)\right) + \tilde{s}_j\left(\text{logit}\left(\widehat{\pi}(\boldsymbol{x}_{ij})\right)\right).$$

Here, $\tilde{s}_j$ denotes the smooth random effect for cluster $j$. The model is fit using `lme4`[38] and splines are added with `rms` package[39]. This approach estimates the calibration per cluster and the variance of the random effects in a single step. As opposed to the random effects in 2MA-C, the MIX-C model takes all observations across the entire spectrum of predicted risk into account when predicting the realized values of the random effects and cluster-specific calibration curves. The main limitation is the computation time needed when the number of clusters is large. The algorithm, additional details on the implementation, and illustration are available in the **Supporting material (Appendix A3, FigureS8), and code for the implementation is available in the OSF repository.**

## 2.7 RESULTS FOR THE CASE EXAMPLE

A visual analysis of the different curves shows that all the approaches present similar results on the case example, in line with the previously obtained results ignoring clustering (**Figure 3**). All plots suggest that ADNEX overestimated risks. Nevertheless, there are important differences in the estimated uncertainty and heterogeneity. Details, visualizations, and variations of each method are shown in **Figures S2-S8. The methods will be available in the CalibrationCurves package from version 2.0.0 onwards.**

# 3 SIMULATION STUDY

The motivating example showcases the applicability of the methods. However, with real data it is not possible to know the true underlying observed proportion. Therefore, we designed a simulation study to explore the performance of the introduced approaches and compare it with cluster ignorant methodologies.

## 3.1 METHODS

The data generating models were based on logistic regression with a random intercept per cluster, hence the formula to obtain the true probabilities was of the form $\text{logit}\left(\pi(x_{ij})\right) = \beta_0 + x_{ij}\beta_1 + u_j$. $\beta_0$ represents the intercept, $\beta_1$ the corresponding effect of covariate $x$ and $u_j$ is the cluster-specific deviation. We first obtain the true models according to a full factorial design where two factors were varied: little vs strong clustering (ICC 5% or 20%) and lower vs higher true area under the receiver operating characteristic curve (AUC;0.75 and 0.9). Event rate was fixed at 30% for the whole population but varied between clusters according to the variance of the random intercept $u_j$. For each data generating mechanism, we generated data for 200 clusters and 2.000.000 patients (10.000 observations per cluster) and a single normally distributed linear predictor ($x$, which can be seen as a combination of several predictors) with mean 0 and variance 1. Random effects ($u_j$) were also normally distributed with variance according to the desired ICC. ICC is calculated as the variance of a null random intercept model divided by the total variance, calculated as the addition of the variances of the null random intercept model and the standard logistic distribution.[40] AUC was controlled by varying the coefficient ($\beta_1$) and the desired event rate was controlled by modifying the intercept $\beta_0$. We obtained 4 superpopulations by trial and error (**Table S1**). The code to obtain the superpopulations is available in the **OSF repository.**

In each of the superpopulations we draw 4 scenarios, inspired by real validation or development studies, varying events per cluster (EPC) (20 vs 200) and number of clusters (5 vs 30). These scenarios refer to the dataset on which hypothetical prediction models are developed or validated. This results in 16 scenarios overall, by combining all values for ICC, AUC, EPC and number of clusters. We then applied the methods for varying development or varying validation sample size.

### 3.1.1 Varying development size

The clusters and patients within each cluster were selected randomly and the number of patients was selected according to the prevalence and the desired EPC ($n_j$ = EPC * 1.15 divided by cluster prevalence, we multiply EPC to ensure that the desired EPC is obtained across all clusters). We developed logistic regression models with restricted cubic splines and 3 knots in these training datasets. Then, we externally validated the calibration of the model in patients from clusters not used for model development in an ideal situation with high number of clusters and high EPC. Therefore, we validated models on a random selection of 100000 observations from a random selection of 30 clusters that were not sampled for model development. On average, each cluster contained 1000 events and 3333 observations, but the number of events varied depending on the cluster specific event rate.

### 3.1.2 Varying validation size

Additionally, we validate the models with varying validation sample size. In this case we fixed the model to be validated to a logistic regression model with restricted cubic splines and 3 knots trained with a sufficiently large sample size according to Riley's criteria (N = 1711) from a cluster with average event rate in each of the superpopulations.[41] Then we validated those models varying EPC (20 vs 200) and number of clusters (5 vs 30) in the validation sample.

We applied the CG-C (grouped and interval), 2MA-C (splines and loess) and MIX-C approaches to the validation sample in both approaches. For comparison, we also obtained a standard flexible logistic calibration model (i.e. ignoring clustering, see section 2.3) using restricted cubic splines and 3 knots. True risks are obtained using the formula in **Table S1**, and the true risks in the cluster with the average effect are obtained by setting the random intercept to 0. These true risks can be used to generate true calibration curves per cluster and for the cluster with the average effect. To numerically compare the deviation of the estimated calibration curve from the true one, we calculate the mean squared calibration error (MSCE) as the mean squared difference between the estimated observed proportions and the true observed proportions (based on the true risks) in the cluster with the average effect (setting $u_j = 0$) over a fixed grid of 100 estimated risks (100 evenly spaced points from 0.01 to 0.99).[23] For CG-C, the grid contains 10 points by definition. The process was repeated 100 times per scenario.

### 3.1.3 Heterogeneity

We evaluated the coverage of the 95% prediction intervals. For each of the three methods, using the same grid of values that we used for the MSCE, we evaluated whether the true cluster-specific risk (including random effects) falls within the prediction interval at each of the grid points. This means that we obtained the true cluster specific probabilities using the formula in **Table S1** and compared for each clustered calibration approach if the prediction interval contained the true values. Coverage was calculated as the proportion over the 200 clusters in the superpopulations.

## 3.2 RESULTS

The median MSCE results for varying validation sample size are displayed in **Table 2** (multiplied by 100) and **Figure 4**. Note that CG-C focuses on 10 points whereas the other approaches focus on a grid of 100 points, such that CG-C and the other approaches are not directly comparable. MIX-C was the best performing approach in all cases. 2MA-C (splines) also performed well in scenarios with high validation sample size. In general, the MSCE was better with high EPC, low ICC and with more clusters in the validation. Standard flexible logistic calibration performed considerably worse in all scenarios, in particular when ICC was high. When evaluated on a big validation sample with varying development size the results showed also that 2MA-C (splines) and MIX-C were the best methods across all scenarios (**Table 3 and Figure S9**).

The prediction interval coverage with varying validation sample was close to nominal with high EPC and low ICC and when using the 2MA-C (splines) approach (**Figure 5**). When EPC was high the coverage was close to nominal in the region where validation data was available but very poor at the tails (i.e. regions with few validation data) except for the MIX-C approach. MIX-C had correct coverage when ICC was high and AUC low, too wide prediction intervals when ICC was low and too narrow PI when ICC and AUC were high. When EPC was low, increasing the number of clusters reduced notably the coverage, especially for the grouped approaches and at the tails. The coverage in the large validation dataset was poor at tails except for MIX-C (**Figure S10**). High AUC and low ICC in development tend to have better PI coverage.

Code and data to reproduce the simulation study and analysis are available in the **OSF repository.**

# 4 SYNTHETIC DATA

The simulation study showed that the methods correctly estimate observed proportions under different logistic DGMs. In the real world, the link between outcome and predictors is often not perfectly defined by a linear or logistic function. Therefore, we aim to test the methods in a more flexible situation using a synthetically generated dataset based on the data used in the motivating example.

## 4.1 METHODS

To generate synthetic data, we use data from the International Ovarian Tumor Analysis (IOTA) consortium that was used to develop prediction models to estimate the risk of malignancy in patients with an ovarian tumor.[9,26] We used the synthpop package in R, which learns the structure of the data and generates a new dataset where individual patient data is masked but the underlying structure is preserved.[42] We generated synthetic data for 1 million individuals for each of the 10 hospitals separately, and used these synthetic patients as cluster-specific populations (retaining the clustered structure of the original dataset). We generated two true models per center: one based on a logistic regression model with splines for continuous variables, and one based on a random forest model with number of randomly selected parameters per split (mtry) set to three and the depth of the individual trees set to ten (minimum node size).[43] Both models were trained on the real data from that center, using the nine ADNEX predictors listed above except type of center. The true models were applied to the 1 million synthetic patients, and the outcome was generated with Bernoulli trials based on the true risks of malignancy from the applied models. This means that the same synthetic patient has two true risks and can have two different outcomes. The code to generate the synthetic data is available in the **OSF repository,** but the original data or the synthetic data are not publicly available. The comparison of the real and synthetic data for quality check in one center is presented in **Figure S11.**

We used the synthetic data to validate the calibration of the published ADNEX model without CA125 in each center.[26] We define 15 scenarios based on the number of centers (2,5,10) and validation data EPC (20,100,200,500,1000). We repeat this 1000 times, by randomly drawing validation datasets. If the number of centers was 2 or 5, the centers were randomly chosen as well. In each repetition, we calculate MSCE for the two true risks. True center-specific curves are obtained by training flexible calibration models (splines with 5 knots) on all 1 million synthetic patients from that center (**Figure 6**). We then compared the center specific estimated observed proportions and the true probabilities per center in a fixed grid of 100 values. The only method that obtains center specific curves borrowing information from other clusters is the MIX-C method and we compare this to standard flexible calibration with LOESS and restricted cubic splines with 3 knots (Stage 1 of 2MA-C).

## 4.2 RESULTS

Median MSCE for the synthetic data analysis with 1000 iterations is shown in **Table 4** and by center in **Figure 7.** MIX-C was the best performing method in all scenarios when the truth was based on a logistic regression, with splines working equally well in 5 scenarios. When the truth was based on a random forest model MIX-C was the best with small validation samples (EPC < 500) and loess when sample size was above 500 events per center. Borrowing information from other clusters is an example of a bias-variance trade-off and leads to improved accuracy for the estimated cluster-specific curves for clusters with modest sample sizes. Additionally, the performance by approach varied considerably between centers (**Figure 7)**. That is, the best approach varied by center.

# 5 DISCUSSION

Calibration of clinical prediction models is crucial since it is related to the usefulness of the recommended clinical decisions provided by the model.[3] Evaluation of calibration performance is best done with flexible calibration plots.[4,20] In this work, we introduced three methods for obtaining flexible calibration plots that account for clustering in the dataset. The first method extended the grouped calibration plot using bivariate random effects meta-analysis (CG-C method), the second method was a two-step approach in which flexible cluster-specific plots were combined through random effects meta-analysis. We generated cluster-specific plots in two ways: using restricted cubic splines (2MA-C splines) or LOESS (2MA-C loess). The third method used a mixed effects logistic calibration model using restricted cubic splines and random intercepts and slopes per cluster (MIX-C). Through a simulation study and a study using synthetic data from patients with an ovarian tumor, we observed that MIX-C and 2MA-C (splines) work best to obtain the calibration plot in the cluster with the average effect (simulation study) and MIX-C for the cluster specific calibration plots (synthetic data). The coverage probability of the prediction interval was suboptimal for all methods and all scenarios with 2MA-C (splines) standing out as the best across scenarios. A disadvantage of 2MA-C (splines) is that the estimated calibration curves per cluster deviated more from the true curves per cluster than those obtained by MIX-C, which uses shrinkage to obtain better cluster-specific calibration curves. While CG-C may work well too to obtain a calibration plot in the cluster with the average effect when using 10 groups based on deciles, it worked poorly when using 10 groups based on equal intervals of the estimated risk. However, the grouped approach tends to depend on the number and types of groups, which is undesirable.[44,45] Obviously, the other methods depend on the level of smoothing of the spline or LOESS, but we automatically selected these parameters based on statistical goodness of fit to reduce the modeler's choices. We recommend using 2MA-C (splines) to estimate the curve with the average effect and the 95% PI and MIX-C for the cluster specific curves, especially when sample size per cluster is limited. In our simulation, 2MA-C (loess) had the worst performance possibly due to the logistic data generation mechanism. In real world the association of outcome and predictors tends to be more complex therefore a more flexible approach could be useful, as shown in the synthetic data with a nonlinear data generating mechanism and large validation sample sizes per cluster. In this work we used the method only for external validation, but the methods also apply to model development with clustered data to explore between-cluster calibration heterogeneity during internal-external validation. We provide ready-to-use R functions to plot the curves and obtain numerical results in the OSF Repository , and they will soon be incorporated into the "`CalibrationCurves`" package in CRAN.[46]

Munoz and colleagues developed similar methodologies to derive calibration plots in large clustered datasets (preprint published after our initial submission).[47] In their work they present four approaches to obtain overall calibration plots. Although their rationale is similar to ours in some cases, the implementation is different. The stacking approach is similar to what we define as flexible curve ignoring clustering (section 2.3), where all the data from different clusters is combined into a single dataset and a calibration model is fitted. Their "One-step meta-analysis" approach corresponds to our MIX-C method where a mixed model is fitted. Their "two step meta-analysis" method is similar to 2MA-C but they also present a variant where instead of fitted probabilities the model parameters are aggregated. Finally, they introduced an approach where the calibration model is a Generalized

Estimating Equation. None of the methodologies include the estimation of prediction intervals. The work is illustrated with an evaluation of a model to diagnose deep vein thrombosis, but no simulation study is included to evaluate the performance of the methodologies or to compare to our results.

The main strength of our methodologies is the inclusion of a heterogeneity measure through the prediction intervals. These intervals indicate, for every estimated risk, the range within which the observed proportion may fall in a new cluster. It is crucial to differentiate the calibration in the cluster with the average effect with the cluster-specific calibration represented in the PIs. Whilst a model can be well calibrated in the cluster with the average effect, this does not necessarily imply that the model is perfectly calibrated in every individual cluster. Statements such as "the model was well calibrated" based on the cluster with the average effect calibration plot should be avoided, or at least accompanied by a clarification that this only applies to the summary curve. Using the interpretation of a calibration curve with the average effect for a specific cluster might lead to incorrect decision making. For example, the curve with average effect suggests overestimation of risks but in some cluster the model might be underestimating them. Although no method provided prediction intervals with correct coverage in all investigated simulation scenarios (**Figure 5 and S10**) they show a more realistic picture than the cluster-ignorant confidence intervals. In general, all methods underestimate the heterogeneity between clusters therefore yielding too narrow prediction intervals, especially for estimated risks far from prevalence and when validation sample size is small. Prediction interval of MIX-C had an odd behaviour where it tended to have too narrow or too wide PIs across the calibration curve. For the rest of approaches, increasing the number of clusters in the validation when events per cluster were low caused the PI to be too narrow (**Figures S12-S16**). More research to fix the suboptimal performance of the prediction intervals is needed where Bayesian based PI derivation could be a solution.[48–50] Another limitation concerning CG-C and 2MA-C is that they present calibration plots based on meta-analysis of independent points. These approaches create the plot joining each pointwise estimate of observed proportion instead of creating a model for the whole plot and therefore are dependent on the number of points. Each of these points represents the calibration in the cluster with the average effect conditional on the estimated risks. We assume that all pointwise estimates form the calibration plot in the cluster with the average effect but in the modelling process each point is independent.

This study is a phase 2 methodological study that focused on the presentation of the methodology and assessments in a limited number of settings.[25] Further research should therefore evaluate the methods in more extended settings and applications. For example, simulations could focus on more complex truths, such as multiple predictors with random slopes for their effects on the outcome, continuous predictors with nonlinear associations with the outcome under study, and prediction models based on flexible machine learning methods such as random forests, boosting approaches, or neural networks. Instead of performing a larger simulation study, we decided to focus on synthetic data based on real data from 10 hospitals on women with an ovarian tumor with the aim of building diagnostic prediction models to estimate the risk of malignancy. This data was used to evaluate an existing prediction model, ADNEX, which was a multinomial logistic regression model using random intercepts and uniform shrinkage of model coefficients.[26] In this synthetic data analysis we focused on the cluster specific performance of MIX-C varying the validation sample. MIX-C showed good results, especially when sample size is limited, while non-clustered approaches generally performed worse.

## AUTHOR CONTRIBUTIONS

Contributions were based on the CRediT taxonomy. Conceptualization: LB, LW, BVC. Funding acquisition: LW, BVC. Project administration: LB. Supervision: LW, BVC. Methodology: LB, LW, BVC, BDCC. Resources: LB. Investigation: LB, LW, BVC. Validation: LB, LW, BVC, BDCC. Data curation: BVC. Software: LB. Formal analysis: LB, LW, BVC. Visualization: LB, LW, BVC. Writing – original draft: LB, LW, BVC. Writing – review & editing: LB, LW, BVC, BDCC. All authors have read, share final responsibility for the decision to submit for publication, and agree to be accountable for all aspects of the work.

## DATA AVAILABILITY

Data and code to reproduce results and figures are available in a public, open access repository (link https://osf.io/aj8ew/). All data relevant to the study are included in the article or uploaded as supplementary information except the motivating example data and the synthetic data due to privacy concerns.

## ACKNOWLEDGMENTS

## FUNDING INFORMATION

This research was supported by the Research Foundation – Flanders (FWO) under grant G097322N with BVC as supervisors. LW and BVC were supported by Internal Funds KU Leuven (grant C24M/20/064). LB is supported by a long stay abroad grant from Research Foundation – Flanders (FWO) under grant V457024N.

## CONFLICT OF INTEREST STATEMENT

The authors declare that there is no conflict of interest.

# REFERENCES

1. Steyerberg EW. *Clinical Prediction Models: A Practical Approach to Development, Validation, and Updating*. Cham: Springer International Publishing; 2019. doi:10.1007/978-3-030-16399-0

2. van Smeden M, Reitsma JB, Riley RD, Collins GS, Moons KG. Clinical prediction models: diagnosis versus prognosis. *Journal of Clinical Epidemiology*. 2021;132:142-145. doi:10.1016/j.jclinepi.2021.01.009

3. Van Calster B, Vickers AJ. Calibration of Risk Prediction Models: Impact on Decision-Analytic Performance. *Med Decis Making*. 2015;35(2):162-169. doi:10.1177/0272989X14547233

4. Van Calster B, Nieboer D, Vergouwe Y, De Cock B, Pencina MJ, Steyerberg EW. A calibration hierarchy for risk models was defined: from utopia to empirical data. *Journal of Clinical Epidemiology*. 2016;74:167-176. doi:10.1016/j.jclinepi.2015.12.005

5. Riley RD, Stewart LA, Tierney JF. Individual Participant Data Meta-Analysis for Healthcare Research. In: *Individual Participant Data Meta-Analysis*. John Wiley & Sons, Ltd; 2021:1-6. doi:10.1002/9781119333784.ch1

6. Wynants L, Kent DM, Timmerman D, Lundquist CM, Van Calster B. Untapped potential of multicenter studies: a review of cardiovascular risk prediction models revealed inappropriate analyses and wide variation in reporting. *Diagn Progn Res*. 2019;3:6. doi:10.1186/s41512-019-0046-9

7. Wessler BS, Paulus J, Lundquist CM, et al. Tufts PACE Clinical Predictive Model Registry: update 1990 through 2015. *Diagnostic and Prognostic Research*. 2017;1(1):20. doi:10.1186/s41512-017-0021-2

8. Debray TPA, Collins GS, Riley RD, et al. Transparent reporting of multivariable prediction models developed or validated using clustered data: TRIPOD-Cluster checklist. *BMJ*. 2023;380:e071018. doi:10.1136/bmj-2022-071018

9. Van Calster B, Valentin L, Froyman W, et al. Validation of models to diagnose ovarian cancer in patients managed surgically or conservatively: multicentre cohort study. *BMJ*. 2020;370:m2614. doi:10.1136/bmj.m2614

10. Gupta RK, Harrison EM, Ho A, et al. Development and validation of the ISARIC 4C Deterioration model for adults hospitalised with COVID-19: a prospective cohort study. *Lancet Respir Med*. 2021;9(4):349-359. doi:10.1016/S2213-2600(20)30559-2

11. Steyerberg EW, Nieboer D, Debray TPA, van Houwelingen HC. Assessment of heterogeneity in an individual participant data meta-analysis of prediction models: An overview and illustration. *Stat Med*. 2019;38(22):4290-4309. doi:10.1002/sim.8296

12. Amsterdam WAC van. A causal viewpoint on prediction model performance under changes in case-mix: discrimination and calibration respond differently for prognosis and diagnosis predictions. September 2024. http://arxiv.org/abs/2409.01444. Accessed October 16, 2024.

13. Wynants L, Vergouwe Y, Van Huffel S, Timmerman D, Van Calster B. Does ignoring clustering in multicenter data influence the performance of prediction models? A simulation study. *Stat Methods Med Res*. 2018;27(6):1723-1736. doi:10.1177/0962280216668555

14. Debray TPA, Moons KGM, Ahmed I, Koffijberg H, Riley RD. A framework for developing, implementing, and evaluating clinical prediction models in an individual participant data meta-analysis. *Statistics in Medicine*. 2013;32(18):3158-3180. doi:10.1002/sim.5732

15. de Jong VMT, Moons KGM, Eijkemans MJC, Riley RD, Debray TPA. Developing more generalizable prediction models from pooled studies and large clustered data sets. *Statistics in Medicine*. 2021;40(15):3533-3559. doi:10.1002/sim.8981

16. Moher D, Schulz KF, Simera I, Altman DG. Guidance for Developers of Health Research Reporting Guidelines. *PLoS Med*. 2010;7(2):e1000217. doi:10.1371/journal.pmed.1000217

17. Snijders TAB, Bosker RJ. *Multilevel Analysis: An Introduction to Basic and Advanced Multilevel Modeling*. London, England, United Kingdom: Sage Publishers; 2012.

18. Finch WH, Bolin JE, Kelley K. *Multilevel Modeling Using R*. 2nd ed. New York: Chapman and Hall/CRC; 2019. doi:10.1201/9781351062268

19. Van Calster B, Collins GS, Vickers AJ, et al. Performance evaluation of predictive AI models to support medical decisions: Overview and guidance. December 2024. doi:10.48550/arXiv.2412.10288

20. Austin PC, Steyerberg EW. Graphical assessment of internal and external calibration of logistic regression models by using loess smoothers. *Statistics in Medicine*. 2014;33(3):517-535. doi:10.1002/sim.5941

21. Austin PC, Harrell FE, Van Klaveren D. Graphical calibration curves and the integrated calibration index (ICI) for survival models. *Statistics in Medicine*. 2020;39(21):2714-2742. doi:10.1002/sim.8570

22. Austin PC, Putter H, Giardiello D, van Klaveren D. Graphical calibration curves and the integrated calibration index (ICI) for competing risk models. *Diagn Progn Res*. 2022;6(1):2. doi:10.1186/s41512-021-00114-6

23. Van Hoorde K, Van Huffel S, Timmerman D, Bourne T, Van Calster B. A spline-based tool to assess and visualize the calibration of multiclass risk predictions. *Journal of Biomedical Informatics*. 2015. doi:10.1016/j.jbi.2014.12.016

24. Debray TPA, Damen JAAG, Snell KIE, et al. A guide to systematic review and meta-analysis of prediction model performance. *BMJ*. 2017;356:i6460. doi:10.1136/bmj.i6460

25. Heinze G, Boulesteix AL, Kammer M, Morris TP, White IR, Initiative the SP of the S. Phases of methodological research in biostatistics—Building the evidence base for new methods. *Biometrical Journal*. 2024;66(1):2200222. doi:10.1002/bimj.202200222

26. Van Calster B, Hoorde KV, Valentin L, et al. Evaluating the risk of ovarian cancer before surgery using the ADNEX model to differentiate between benign, borderline, early and advanced stage invasive, and secondary metastatic tumours: prospective multicentre diagnostic study. *BMJ*. 2014;349:g5920. doi:10.1136/bmj.g5920

27. Timmerman D, Ledger A, Bourne T, et al. IOTA Phase 1: development of models to distinguish between a benign and malignant adnexal tumor before surgery. February 2024. doi:10.48804/HDFDWI

28. Kaijser J. Towards an evidence-based approach for diagnosis and management of adnexal masses: findings of the International Ovarian Tumour Analysis (IOTA) studies. *Facts Views Vis Obgyn*. 2015;7(1):42-59.

29. Harrell , FE. *Regression Modeling Strategies: With Applications to Linear Models, Logistic and Ordinal Regression, and Survival Analysis*. Cham: Springer International Publishing; 2015. doi:10.1007/978-3-319-19425-7

30. Dimitriadis T, Dümbgen L, Henzi A, Puke M, Ziegel J. Honest calibration assessment for binary outcome predictions. *Biometrika*. 2023;110(3):663-680. doi:10.1093/biomet/asac068

31. Campo BDC. Towards reliable predictive analytics: a generalized calibration framework. September 2023. doi:10.48550/arXiv.2309.08559

32. On behalf of Topic Group 'Evaluating diagnostic tests and prediction models' of the STRATOS initiative, Van Calster B, McLernon DJ, van Smeden M, Wynants L, Steyerberg EW. Calibration: the Achilles heel of predictive analytics. *BMC Med*. 2019;17(1):230. doi:10.1186/s12916-019-1466-7

33. Cox DR. Two Further Applications of a Model for Binary Regression. *Biometrika*. 1958;45(3/4):562-565. doi:10.2307/2333203

34. Wickham H, Averick M, Bryan J, et al. Welcome to the tidyverse. *Journal of Open Source Software*. 2019;4(43):1686. doi:10.21105/joss.01686

35. R Core Team. *R: A Language and Environment for Statistical Computing*. Vienna, Austria: R Foundation for Statistical Computing; 2024. https://www.R-project.org/.

36. Viechtbauer W. Conducting Meta-Analyses in R with the metafor Package. *Journal of Statistical Software*. 2010;36:1-48. doi:10.18637/jss.v036.i03

37. Barreñada L, Wynants L, Calster B van. Clustered Flexible Calibration Plots For Binary Outcomes Using Random Effects Modeling. July 2024. https://osf.io/erju9/. Accessed March 11, 2025.

38. Bates D, Mächler M, Bolker B, Walker S. Fitting Linear Mixed-Effects Models Using **lme4**. *J Stat Soft*. 2015;67(1). doi:10.18637/jss.v067.i01

39. Harrell Jr FE. rms: Regression Modeling Strategies. September 2009:7.0-0. doi:10.32614/CRAN.package.rms

40. Hosmer DW, Lemeshow S, Sturdivant RX. *Applied Logistic Regression*. 1st ed. Wiley; 2013. doi:10.1002/9781118548387

41. Riley RD, Debray TPA, Collins GS, et al. Minimum sample size for external validation of a clinical prediction model with a binary outcome. *Statistics in Medicine*. 2021;40(19):4230-4251. doi:10.1002/sim.9025

42. Nowok B, Raab GM, Dibben C. synthpop: Bespoke Creation of Synthetic Data in R. *Journal of Statistical Software*. 2016;74:1-26. doi:10.18637/jss.v074.i11

43. Wright MN, Ziegler A. ranger: A Fast Implementation of Random Forests for High Dimensional Data in C++ and R. *Journal of Statistical Software*. 2017;77:1-17. doi:10.18637/jss.v077.i01

44. Van Calster B, Collins GS, Vickers AJ, et al. Performance evaluation of predictive AI models to support medical decisions: Overview and guidance. December 2024. doi:10.48550/arXiv.2412.10288

45. Arrieta-Ibarra I, Gujral P, Tannen J, Tygert M, Xu C. Metrics of calibration for probabilistic predictions. June 2022. doi:10.48550/arXiv.2205.09680

46. De Cock B, Nieboer D, Van Calster B, Steyerberg EW, Vergouwe Y. The CalibrationCurves package: assessing the agreement between observed outcomes and predictions. 2023. doi:10.32614/CRAN.package.CalibrationCurves

47. Munoz J, Moons KGM, Debray TPA, Jong VMT de. Deriving calibration plots in large clustered datasets. June 2025. doi:10.21203/rs.3.rs-6740452/v1

48. Partlett C, Riley RD. Random effects meta-analysis: Coverage performance of 95% confidence and prediction intervals following REML estimation. *Stat Med*. 2017;36(2):301-317. doi:10.1002/sim.7140

49. Debray TP, Damen JA, Riley RD, et al. A framework for meta-analysis of prediction model studies with binary and time-to-event outcomes. *Stat Methods Med Res*. 2019;28(9):2768-2786. doi:10.1177/0962280218785504

50. Higgins JPT, Thompson SG, Spiegelhalter DJ. A re-evaluation of random-effects meta-analysis. *J R Stat Soc Ser A Stat Soc*. 2009;172(1):137-159. doi:10.1111/j.1467-985X.2008.00552.x

# Figures

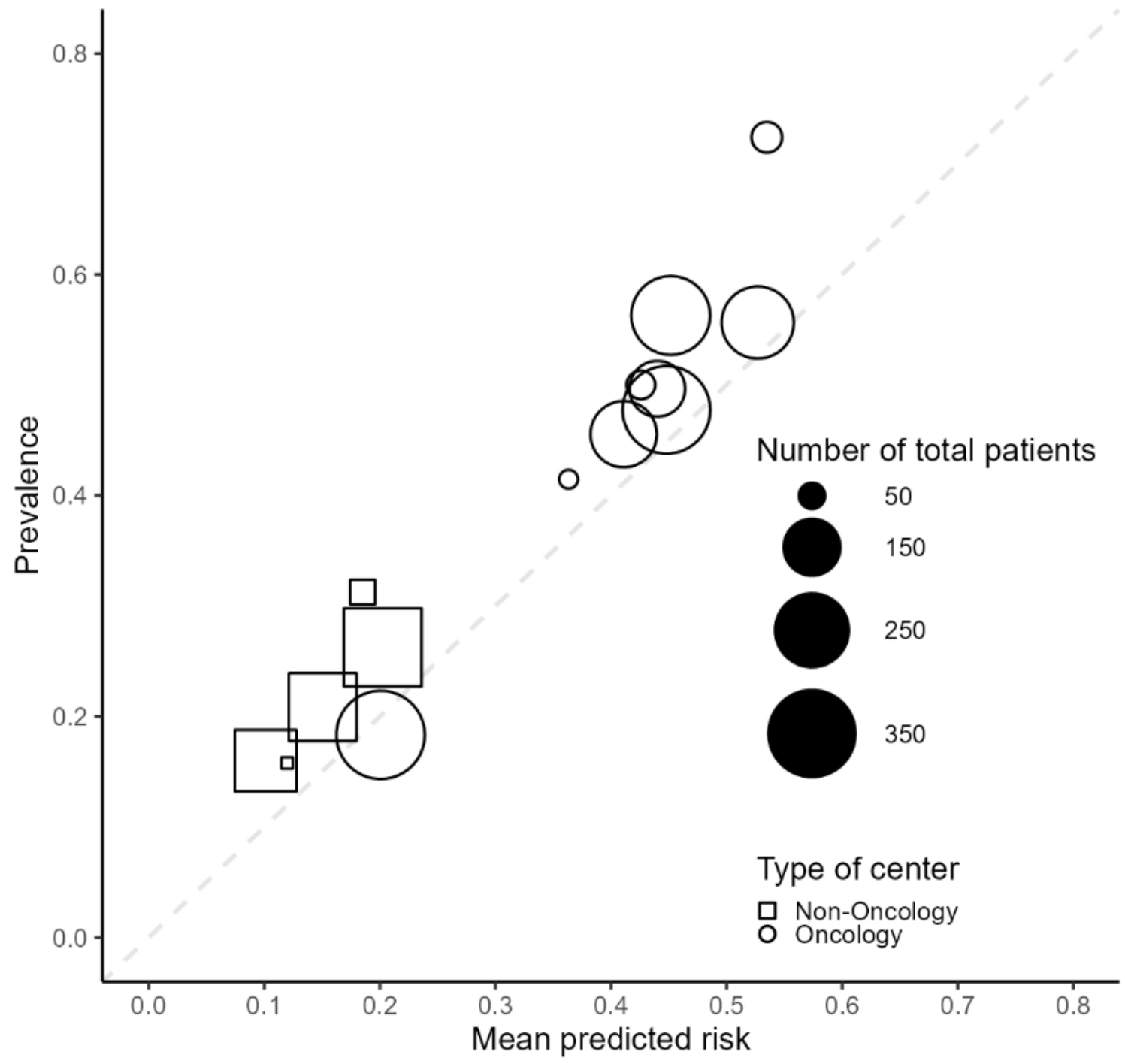

**Figure 1. Prevalence and mean predicted ADNEX risk by center across the 14 centers in the dataset. Dashed diagonal line indicates perfect calibration.**

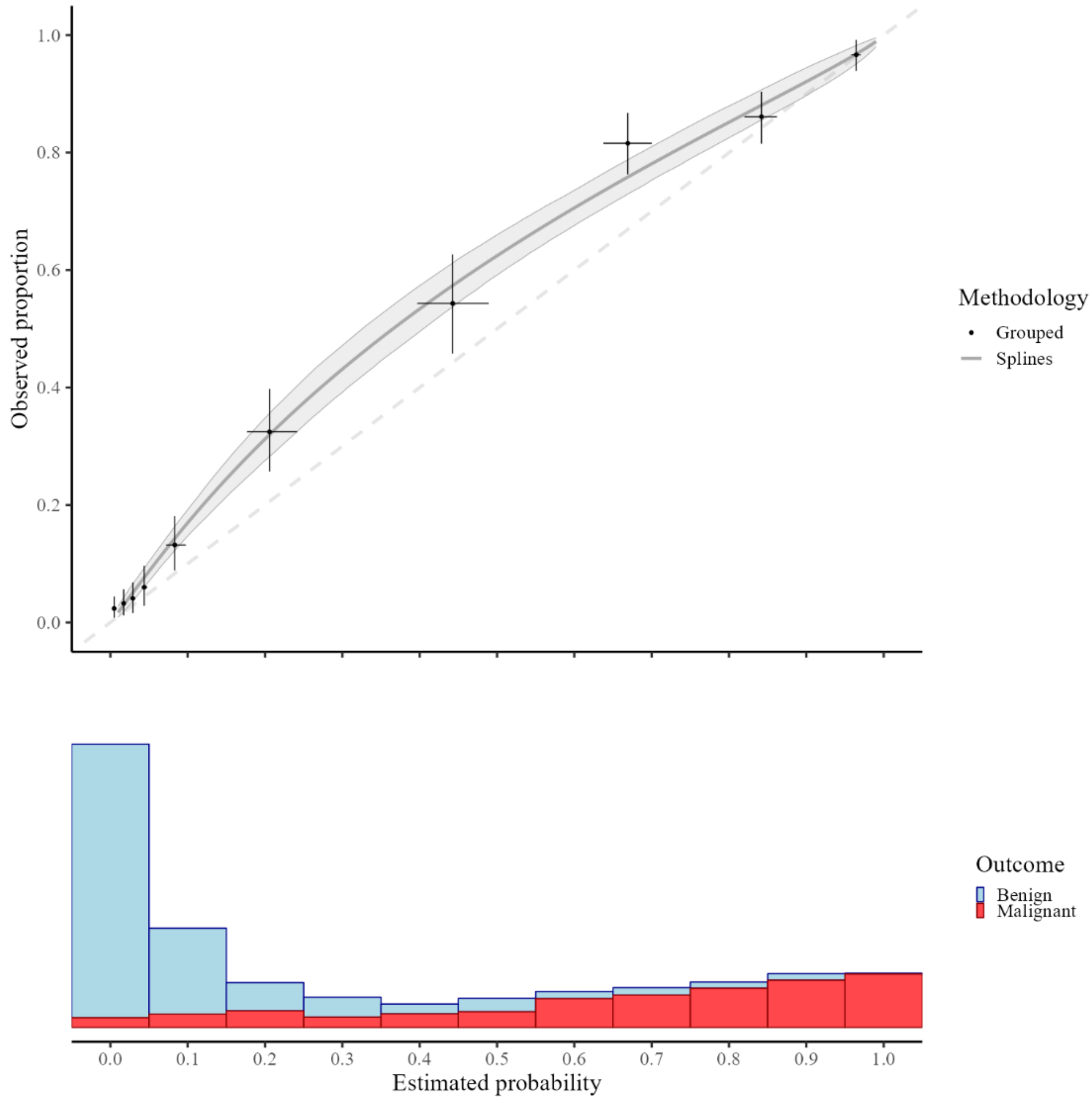

**Figure 2. Traditional flexible calibration curves for the ADNEX model in the motivating example. Observed proportion is estimated with a logistic model with restricted cubic splines to model nonlinear effects and estimated risks are grouped in 10 groups. Confidence intervals are shown for 1000 bootstraps with a shaded area for splines and a + for grouped calibration. Dashed diagonal line indicates perfect calibration.**

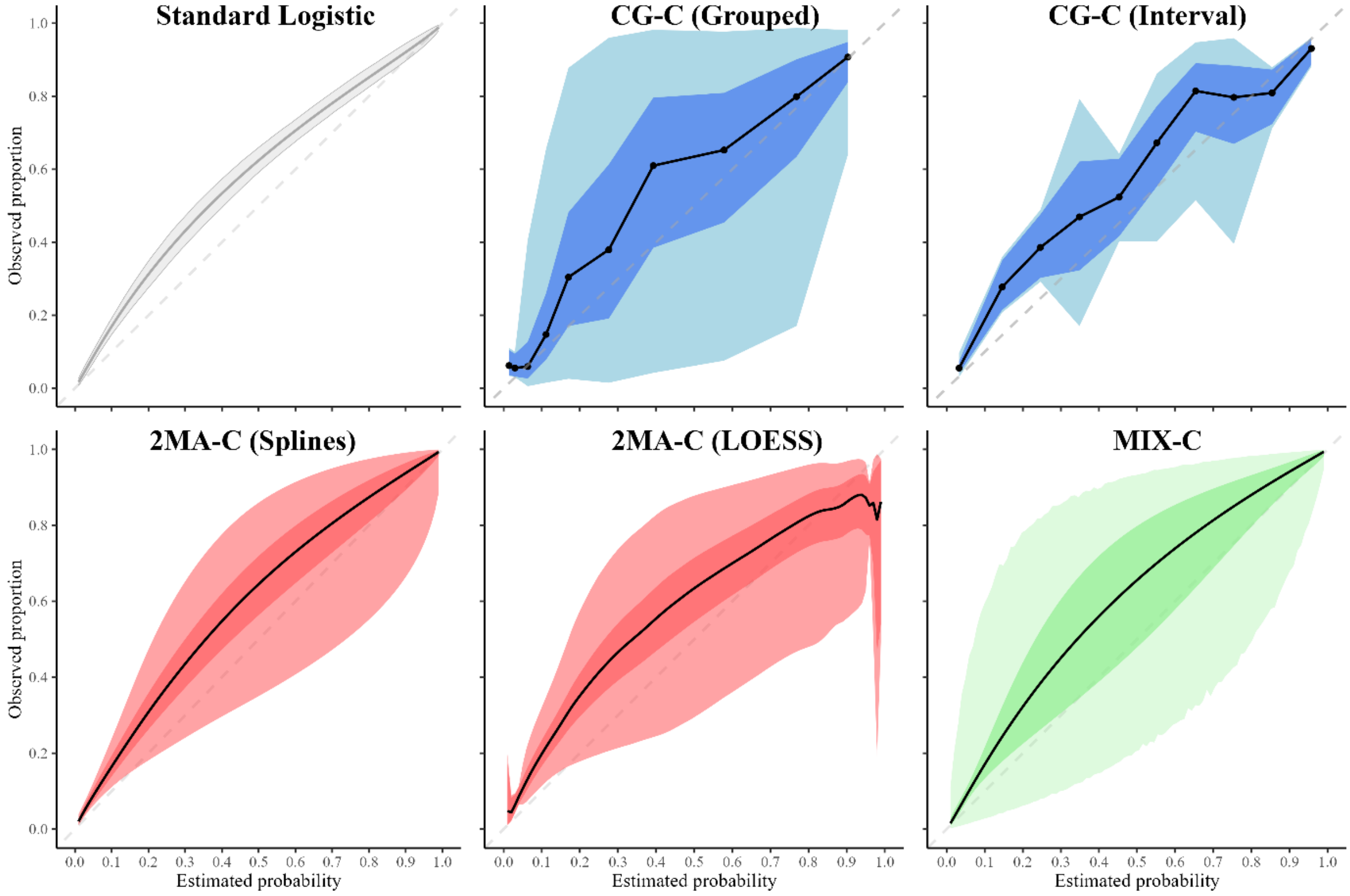

**Figure 3. Comparison of the standard logistic regression with splines and the 3 introduced methodologies with confidence (bright shaded) and prediction intervals (light shaded). Number of quantiles for CG-C were 10, 2MA-C fitted center specific curves with splines or LOESS and MIX-C used random intercept and slopes with restricted cubic splines and 3 knots. Dashed diagonal line indicates perfect calibration.**

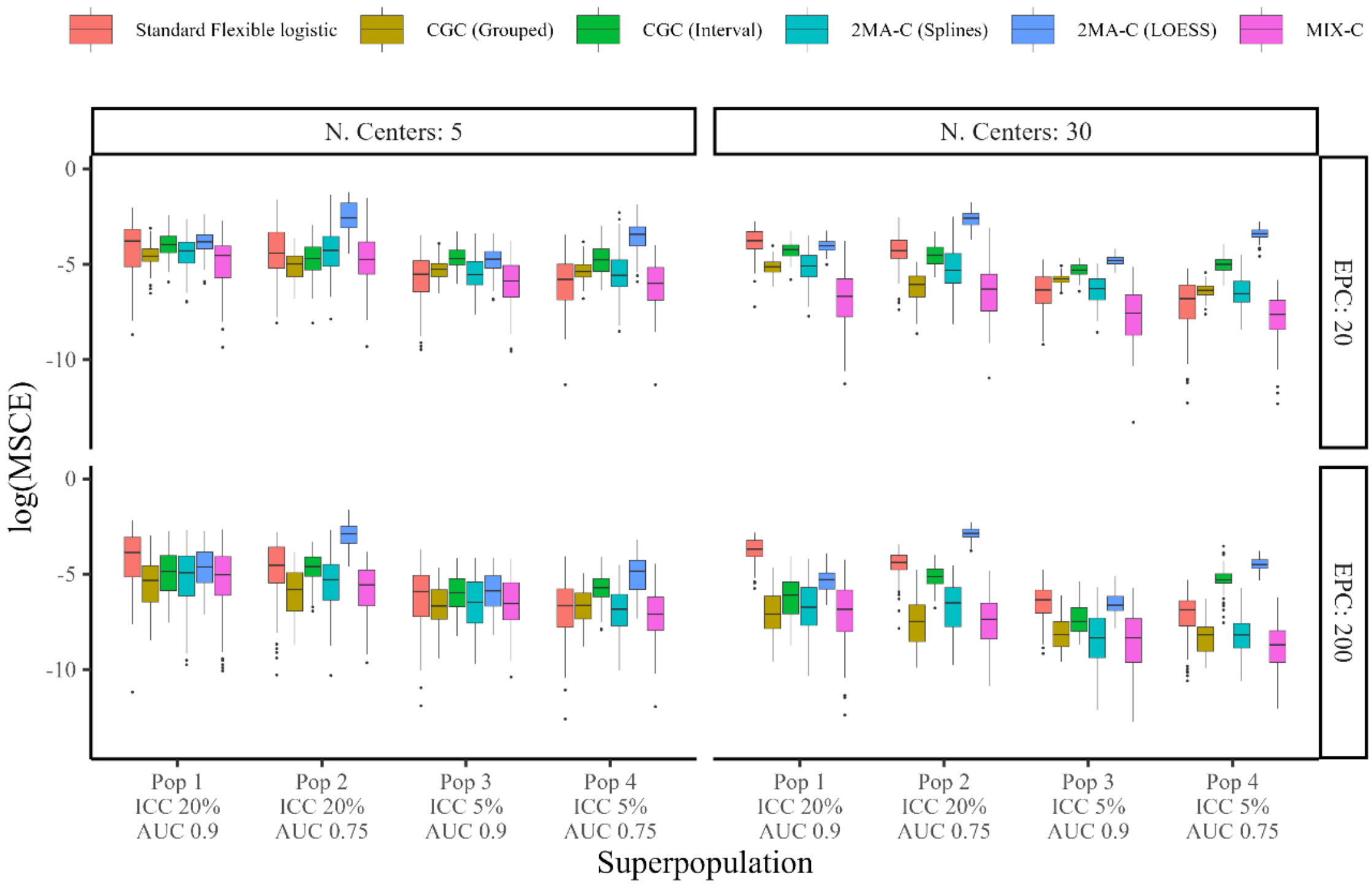

**Figure 4. Boxplots of mean squared calibration error (log) for fixed prediction models varying validation events per center (EPC) and number of centers in the validation.**

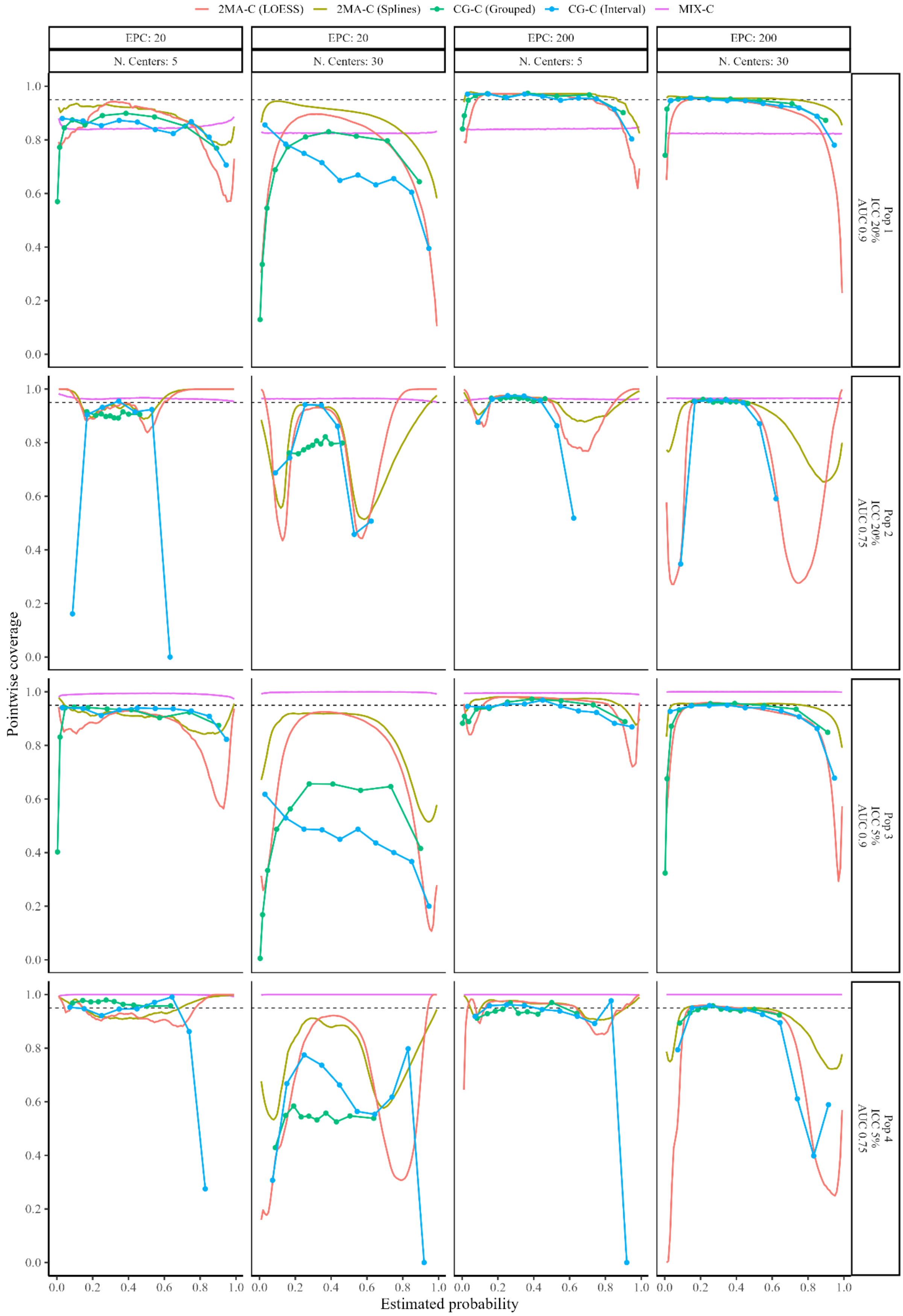

**Figure 5. Pointwise prediction interval coverage with varying validation sample size. The model validated is the same in each superpopulation and it was trained from a center with average event rate and with adequate sample size. Black dotted line indicates nominal coverage (95%).**

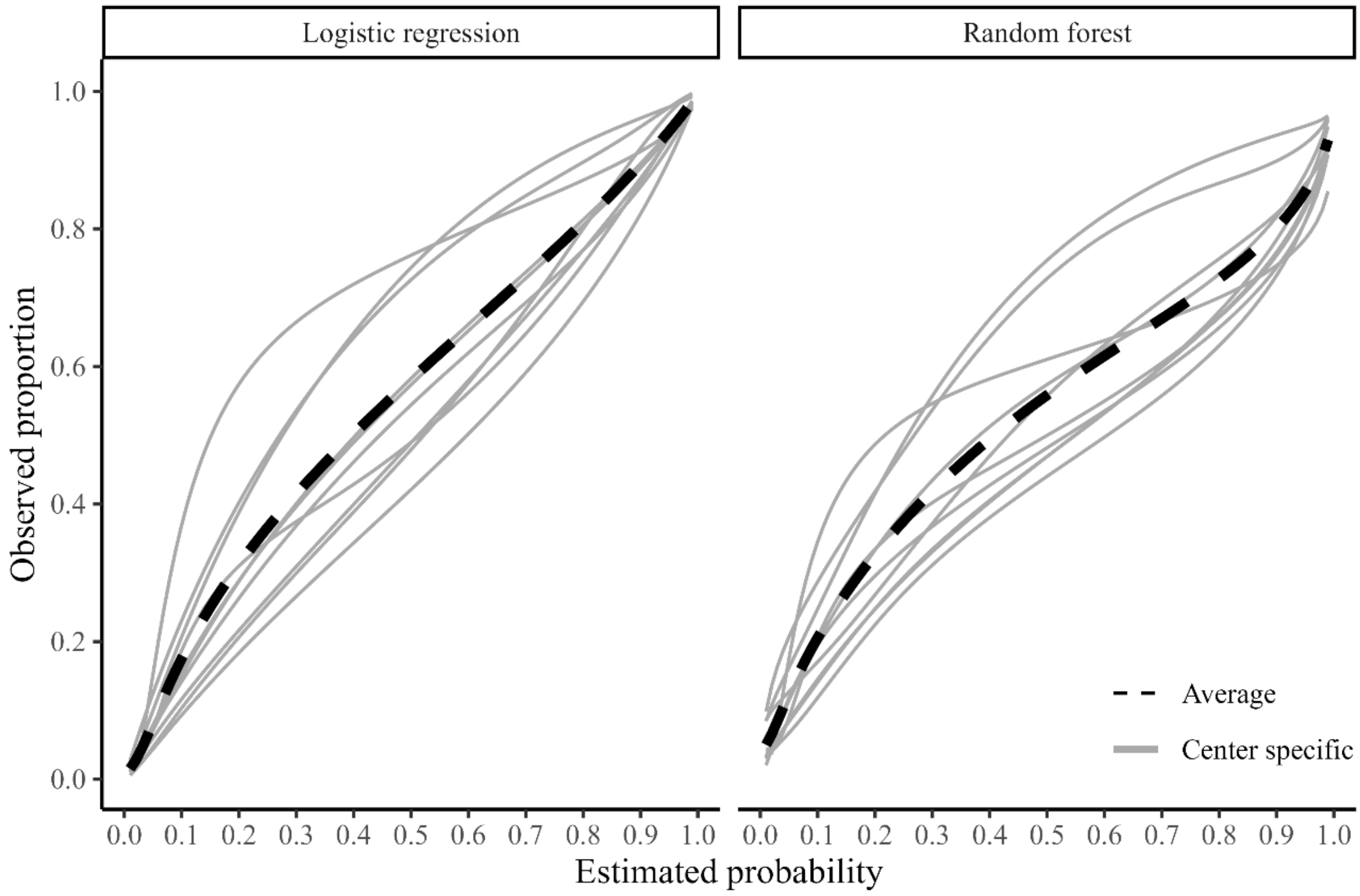

**Figure 6. Center specific (grey) and average true calibration plots for the synthetic data with 1000000 observations per center.**

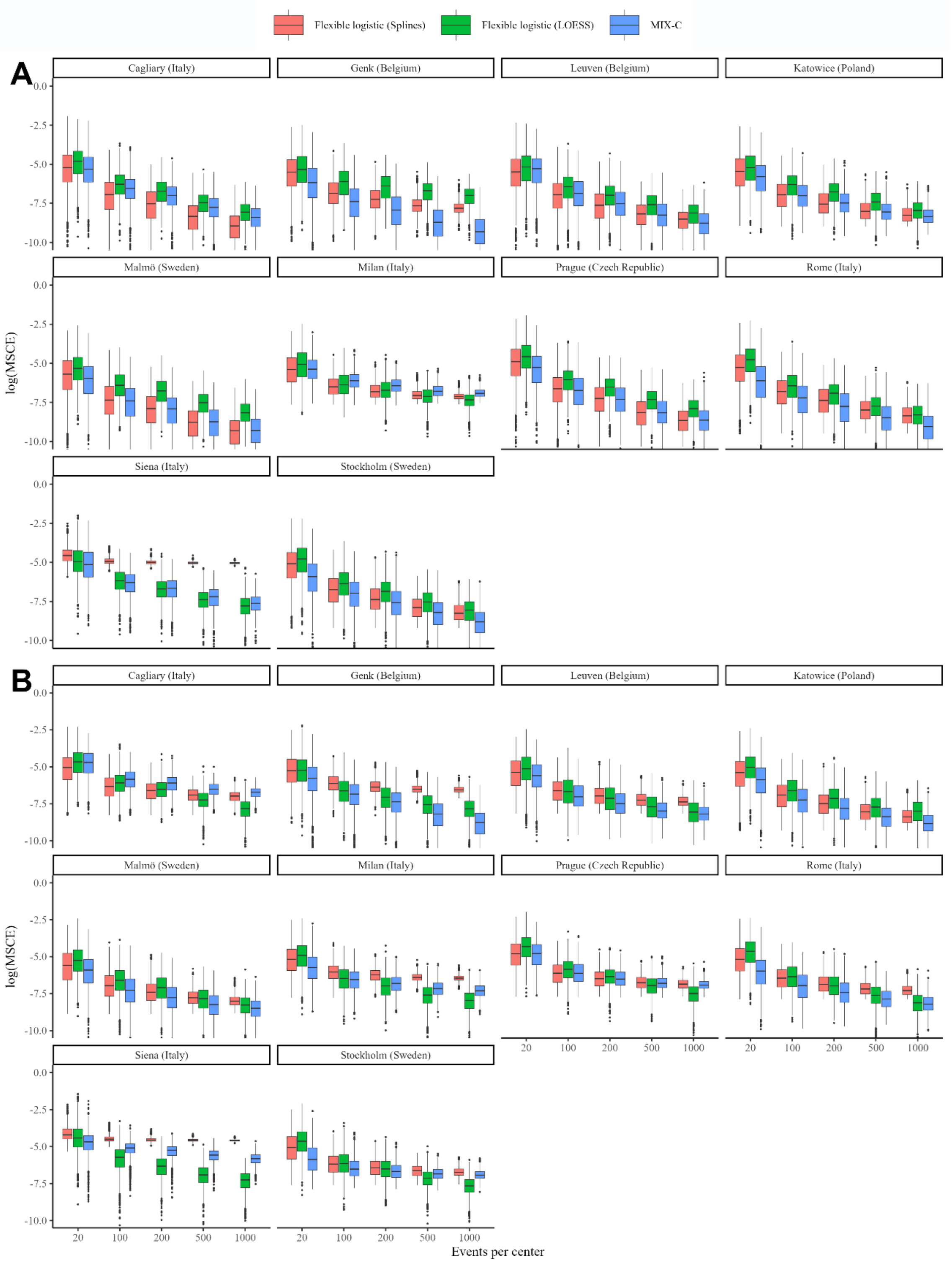

**Figure 7. Center specific results of MSCE (multiplied by 100) by the number of events per center for the logistic truth (A) and the random forest (B) truth.**

# Tables

| Method | Estimation of observed proportion | Strengths | Limitations | Implementation | Illustration |
|---|---|---|---|---|---|
| **CG-C** | **Grouped**: Bivariate random effects meta-analysis of logit-transformed mean estimated risk and event rate by quantile per cluster.<br><br>**Interval**: Bivariate random effects meta-analysis of logit-transformed mean estimated risk and event rate by estimated risk interval. | Model agnostic<br><br>Pointwise confidence and prediction intervals.<br><br>All clusters have the same number of groups | Computation time<br><br>Groups can contain observations with very different estimated risks within and between clusters (Grouped version).<br><br>Clusters may not have the same number of groups (e.g. risk intervals without observations). (Interval version)<br><br>Curves depend on number of groups | CGC(method = "grouped")<br><br>CGC(method = "interval") | Figures S2-3<br><br>Figures S4-S5 |
| **2MA-C** | Random effects meta-analysis of estimated smooth observed proportion by cluster<br><br>**Splines**: Recommended when clusters are small<br><br>**LOESS**: More flexible but can fail with small clusters. | Pointwise confidence and prediction intervals. | Computation time<br><br>Curve dependent on the smoother used in the cluster-specific models. | MAC2(method_choice = "splines")<br><br>MAC2(method_choice = "loess") | Figure S6 (Splines)<br><br>Figure S7 (LOESS) |
| **MIX-C** | Logistic generalized linear mixed model with restricted cubic splines. | Curvewise confidence and prediction intervals.<br><br>Provides also shrunken curves per center. | Computation time | MIXC(model = "slope") | Figure S8(slope) |

**Table 1. Overview of introduced methodologies for creating flexible calibration curves accounting for clustering.**

**Table 2. Median (IQR) squared difference between true average probabilities and estimated observed proportion (MSCE) with logistic calibration, CG-C (10 groups) ,2MA-C and MIX-C method for a logistic model varying validation events per center and number of centers. MSCE is multiplied by 100. The lower the number, the closer the estimated summary calibration curve is to the true calibration curve in a cluster with an average effect. Rounded to two decimals.**

| EPC | N. Cent | ICC | AUC | Standard Flexible Logistic | CG-C group* | CG-C interval* | 2MA-C Splines | 2MA-C LOESS | MIX-C |
|---|---|---|---|---|---|---|---|---|---|
| 20 | 5 | 0.05 | 0.75 | 0.31 (0.10-0.68) | 0.46 (0.34-0.66) | 0.85 (0.47-1.51) | 0.38 (0.21-0.85) | 3.26 (1.78-4.79) | **0.25 (0.10-0.57)** |
| 20 | 5 | 0.05 | 0.9 | 0.40 (0.16-0.81) | 0.52 (0.35-0.70) | 0.90 (0.66-1.42) | 0.39 (0.23-0.78) | 0.89 (0.55-1.32) | **0.28 (0.12-0.64)** |
| 20 | 5 | 0.2 | 0.75 | 1.22 (0.55-3.63) | 0.69 (0.35-1.04) | 0.91 (0.50-1.67) | 1.40 (0.61-2.91) | 7.67 (4.59-16.77) | **0.87 (0.40-2.18)** |
| 20 | 5 | 0.2 | 0.9 | 2.29 (0.60-4.26) | 1.03 (0.79-1.51) | 1.89 (1.24-2.94) | 1.36 (0.72-2.15) | 2.20 (1.52-3.17) | **1.07 (0.33-1.77)** |
| 20 | 30 | 0.05 | 0.75 | 0.11 (0.04-0.23) | 0.17 (0.14-0.22) | 0.67 (0.50-0.87) | 0.14 (0.09-0.28) | 3.36 (2.82-4.08) | **0.05 (0.02-0.10)** |
| 20 | 30 | 0.05 | 0.9 | 0.18 (0.09-0.35) | 0.31 (0.26-0.36) | 0.50 (0.40-0.65) | 0.19 (0.10-0.32) | 0.82 (0.69-0.97) | **0.05 (0.02-0.14)** |
| 20 | 30 | 0.2 | 0.75 | 1.38 (0.93-2.37) | 0.23 (0.12-0.36) | 1.09 (0.69-1.65) | 0.49 (0.25-1.18) | 7.59 (5.36-9.76) | **0.18 (0.06-0.40)** |
| 20 | 30 | 0.2 | 0.9 | 2.32 (1.53-3.76) | 0.59 (0.45-0.76) | 1.46 (1.05-1.85) | 0.62 (0.35-1.07) | 1.78 (1.41-2.22) | **0.13 (0.04-0.31)** |
| 200 | 5 | 0.05 | 0.75 | 0.13 (0.04-0.31) | 0.13 (0.06-0.25) | 0.33 (0.21-0.54) | 0.11 (0.05-0.23) | 0.79 (0.30-1.40) | **0.08 (0.04-0.20)** |
| 200 | 5 | 0.05 | 0.9 | 0.27 (0.07-0.64) | 0.13 (0.06-0.30) | 0.25 (0.12-0.53) | **0.15 (0.05-0.44)** | 0.28 (0.13-0.62) | **0.15 (0.06-0.43)** |
| 200 | 5 | 0.2 | 0.75 | 1.07 (0.43-2.80) | 0.30 (0.10-0.73) | 1.01 (0.60-1.67) | 0.51 (0.17-1.13) | 5.56 (3.41-8.37) | **0.39 (0.13-0.84)** |
| 200 | 5 | 0.2 | 0.9 | 2.10 (0.58-4.67) | 0.49 (0.16-1.06) | 0.78 (0.29-1.79) | 0.73 (0.22-1.76) | 0.98 (0.43-2.16) | **0.66 (0.22-1.69)** |
| 200 | 30 | 0.05 | 0.75 | 0.10 (0.04-0.17) | 0.03 (0.01-0.04) | 0.50 (0.42-0.69) | 0.03 (0.01-0.05) | 1.12 (0.91-1.43) | **0.02 (0.01-0.03)** |
| 200 | 30 | 0.05 | 0.9 | 0.18 (0.09-0.29) | 0.03 (0.02-0.06) | 0.06 (0.03-0.12) | **0.02 (0.01-0.07)** | 0.13 (0.10-0.21) | **0.02 (0.01-0.07)** |
| 200 | 30 | 0.2 | 0.75 | 1.25 (0.85-1.83) | 0.06 (0.02-0.14) | 0.59 (0.41-0.88) | 0.15 (0.04-0.33) | 5.73 (4.70-7.18) | **0.06 (0.02-0.15)** |
| 200 | 30 | 0.2 | 0.75 | 2.50 (1.68-3.97) | 0.08 (0.04-0.21) | 0.23 (0.08-0.45) | 0.12 (0.05-0.34) | 0.51 (0.31-0.72) | **0.11 (0.03-0.29)** |

*Deciles is calculated only with 10 points instead of 100. In green the best working model(s) for estimating observed proportions excluding CG-C methods.
EPC: Event per center in **validation** sample
N.Cent: Number of centers in **validation** sample

**Table 3. Median (IQR) squared difference between true average probabilities and estimated observed proportion (MSCE) with logistic calibration, CG-C (10 groups) ,2MA-C and MIX-C method for a logistic model varying training sample events per center and number of centers. MSCE is multiplied by 100. The lower the number, the closer the estimated summary calibration curve is to the true calibration curve in a cluster with an average effect. Rounded to two decimals.**

| EPC | N. Cent | ICC | AUC | Standard Flexible Logistic | CG-C group* | CG-C interval* | 2MA-C Splines | 2MA-C LOESS | MIX-C |
|---|---|---|---|---|---|---|---|---|---|
| 20 | 5 | 0.05 | 0.75 | 0.03 (0.02-0.06) | 0.01 (0.00-0.03) | 0.51 (0.27-0.70) | **0.01 (0.00-0.03)** | 0.54 (0.16-1.45) | 0.02 (0.01-0.04) |
| 20 | 5 | 0.05 | 0.9 | 0.05 (0.04-0.09) | 0.02 (0.01-0.04) | 0.03 (0.01-0.06) | **0.03 (0.01-0.05)** | 0.06 (0.03-0.10) | **0.03 (0.01-0.06)** |
| 20 | 5 | 0.2 | 0.75 | 0.27 (0.13-0.80) | 0.06 (0.02-0.16) | 1.13 (0.56-2.07) | **0.06 (0.02-0.17)** | 4.54 (3.13-8.44) | 0.12 (0.03-0.61) |
| 20 | 5 | 0.2 | 0.9 | 0.36 (0.27-0.46) | 0.05 (0.02-0.18) | 0.17 (0.06-0.36) | **0.06 (0.01-0.20)** | 0.46 (0.29-0.87) | **0.06 (0.02-0.20)** |
| 20 | 30 | 0.05 | 0.75 | **0.02 (0.01-0.04)** | 0.02 (0.00-0.04) | 0.47 (0.33-0.59) | **0.02 (0.00-0.03)** | 0.40 (0.20-0.72) | **0.02 (0.00-0.03)** |
| 20 | 30 | 0.05 | 0.9 | 0.05 (0.03-0.09) | 0.02 (0.01-0.04) | 0.02 (0.01-0.07) | **0.02 (0.01-0.07)** | 0.05 (0.03-0.09) | **0.02 (0.01-0.06)** |
| 20 | 30 | 0.2 | 0.75 | 0.16 (0.10-0.24) | 0.04 (0.01-0.14) | 1.15 (0.72-1.75) | **0.04 (0.02-0.11)** | 6.06 (4.61-8.15) | 0.05 (0.02-0.13) |
| 20 | 30 | 0.2 | 0.9 | 0.36 (0.27-0.44) | 0.06 (0.02-0.15) | 0.18 (0.08-0.32) | **0.06 (0.02-0.17)** | 0.63 (0.46-0.91) | **0.06 (0.02-0.16)** |
| 200 | 5 | 0.05 | 0.75 | 0.02 (0.01-0.03) | 0.01 (0.00-0.03) | 0.42 (0.34-0.58) | **0.01 (0.00-0.02)** | 0.41 (0.24-0.61) | **0.01 (0.00-0.02)** |
| 200 | 5 | 0.05 | 0.9 | 0.05 (0.03-0.08) | 0.02 (0.01-0.04) | 0.03 (0.01-0.06) | **0.02 (0.01-0.05)** | 0.05 (0.03-0.10) | **0.02 (0.01-0.05)** |
| 200 | 5 | 0.2 | 0.75 | 0.14 (0.10-0.21) | 0.05 (0.02-0.13) | 1.23 (0.64-2.03) | **0.05 (0.02-0.11)** | 6.55 (4.45-7.90) | **0.05 (0.02-0.11)** |
| 200 | 5 | 0.2 | 0.9 | 0.34 (0.24-0.46) | 0.06 (0.03-0.18) | 0.13 (0.06-0.36) | **0.06 (0.03-0.21)** | 0.46 (0.26-0.90) | **0.06 (0.03-0.19)** |
| 200 | 30 | 0.05 | 0.75 | 0.02 (0.01-0.03) | 0.01 (0.00-0.03) | 0.44 (0.38-0.54) | **0.01 (0.00-0.03)** | 0.41 (0.22-0.68) | **0.01 (0.00-0.03)** |
| 200 | 30 | 0.05 | 0.9 | 0.05 (0.03-0.08) | 0.02 (0.00-0.04) | 0.02 (0.01-0.07) | **0.02 (0.01-0.07)** | 0.04 (0.03-0.10) | **0.02 (0.00-0.07)** |
| 200 | 30 | 0.2 | 0.75 | 0.13 (0.09-0.19) | 0.05 (0.02-0.14) | 1.17 (0.62-2.05) | 0.05 (0.02-0.12) | 6.71 (5.57-7.72) | **0.04 (0.02-0.10)** |
| 200 | 30 | 0.2 | 0.9 | 0.31 (0.24-0.51) | 0.07 (0.02-0.18) | 0.16 (0.09-0.32) | 0.08 (0.01-0.20) | 0.62 (0.46-0.85) | **0.07 (0.01-0.19)** |

*Deciles is calculated only with 10 points instead of 100. In green the best working model(s) for estimating observed proportions excluding CG-C method
EPC: Event per center in **training** sample
N.Cent: Number of centers in **training** sample

**Table 4. Median MSCE for the synthetic data study for center specific results. MSCE is presented multiplied by 100. Note that the center specific true curves are based on a flexible logistic model with restricted cubic splines. The lower the number, the closer the estimated center-specific calibration curve is to the true calibration curve in each cluster. Rounded to two decimals.**

| | | Logistic regression truth | | | Random forest truth | | |
|---|---|---|---|---|---|---|---|
| EPC | N. centers | Flexible logistic (Splines) | Flexible logistic (LOESS) | MIX-C | Flexible logistic (Splines) | Flexible logistic (LOESS) | MIX-C |
| 20 | 2 | 0.54 (0.21-1.11) | 0.66 (0.29-1.37) | **0.52 (0.23-1.01)** | 0.65 (0.28-1.38) | 0.81 (0.39-1.63) | **0.58 (0.25-1.19)** |
| 20 | 5 | 0.57 (0.22-1.19) | 0.68 (0.30-1.43) | **0.39 (0.15-0.85)** | 0.66 (0.26-1.31) | 0.83 (0.39-1.63) | **0.44 (0.19-0.96)** |
| 20 | 10 | 0.55 (0.22-1.13) | 0.67 (0.3-1.41) | **0.33 (0.13-0.72)** | 0.64 (0.27-1.31) | 0.81 (0.38-1.62) | **0.38 (0.16-0.85)** |
| 100 | 2 | **0.13 (0.06-0.29)** | 0.19 (0.10-0.36) | **0.13 (0.06-0.27)** | 0.19 (0.10-0.38) | 0.19 (0.09-0.35) | **0.17 (0.08-0.35)** |
| 100 | 5 | 0.13 (0.05-0.28) | 0.18 (0.09-0.34) | **0.12 (0.05-0.23)** | 0.19 (0.10-0.39) | 0.19 (0.09-0.36) | **0.15 (0.07-0.32)** |
| 100 | 10 | 0.13 (0.06-0.28) | 0.18 (0.09-0.34) | **0.11 (0.04-0.22)** | 0.20 (0.10-0.38) | 0.18 (0.09-0.35) | **0.14 (0.06-0.30)** |
| 200 | 2 | **0.08 (0.03-0.15)** | 0.11 (0.06-0.21) | **0.08 (0.03-0.15)** | 0.14 (0.08-0.25) | 0.12 (0.06-0.20) | **0.11 (0.05-0.23)** |
| 200 | 5 | **0.07 (0.03-0.16)** | 0.12 (0.07-0.21) | **0.07 (0.03-0.14)** | 0.14 (0.08-0.25) | 0.12 (0.06-0.20) | **0.10 (0.05-0.21)** |
| 200 | 10 | **0.07 (0.03-0.16)** | 0.12 (0.06-0.21) | **0.07 (0.03-0.13)** | 0.14 (0.07-0.25) | 0.11 (0.06-0.20) | **0.10 (0.04-0.20)** |
| 500 | 2 | **0.04 (0.02-0.08)** | 0.06 (0.03-0.11) | **0.04 (0.02-0.08)** | 0.11 (0.06-0.18) | **0.06 (0.03-0.11)** | 0.07 (0.03-0.13) |
| 500 | 5 | 0.04 (0.02-0.08) | 0.06 (0.03-0.11) | **0.03 (0.02-0.07)** | 0.11 (0.06-0.17) | **0.06 (0.03-0.11)** | 0.07 (0.03-0.13) |
| 500 | 10 | 0.04 (0.02-0.08) | 0.06 (0.03-0.11) | **0.03 (0.01-0.07)** | 0.11 (0.06-0.17) | **0.06 (0.03-0.11)** | 0.07 (0.03-0.13) |
| 1000 | 2 | 0.03 (0.01-0.06) | 0.04 (0.02-0.06) | **0.02 (0.01-0.05)** | 0.10 (0.05-0.15) | **0.04 (0.02-0.06)** | 0.05 (0.02-0.11) |
| 1000 | 5 | 0.03 (0.01-0.06) | 0.04 (0.02-0.07) | **0.02 (0.01-0.04)** | 0.10 (0.05-0.15) | **0.04 (0.02-0.07)** | 0.06 (0.02-0.11) |
| 1000 | 10 | 0.03 (0.01-0.06) | 0.04 (0.02-0.07) | **0.02 (0.01-0.04)** | 0.10 (0.05-0.15) | **0.04 (0.02-0.07)** | 0.06 (0.02-0.11) |

# CLUSTERED FLEXIBLE CALIBRATION PLOTS FOR BINARY OUTCOMES USING RANDOM EFFECTS MODELING

Lasai Barreñada, Bavo D.C. Campo, Laure Wynants, Ben Van Calster

# SUPPLEMENTARY MATERIAL

**Corresponding author**

Ben Van Calster

KU Leuven, Department of Development and Regeneration

Herestraat 49 box 805

3000 Leuven

Belgium

Ben.vancalster@kuleuven.be

# APPENDIX

## A1. Clustered group calibration (CG-C)

Clustered group calibration can be seen as an extension of the traditional grouped calibration or binning calibration, where the data is split into equally sized groups based on the distribution of estimated risks, and the calibration curve shows for each group the estimated prevalence of the event on the y-axis and the mean estimated risks on the x-axis.

**CG-C (grouped)**

1. For each cluster $j = (1, \ldots, J)$, we group the estimated probabilities into $Q$ quantiles. Quantiles typically differ by cluster.
2. For each cluster $j$ within quantile $q$, we calculate the mean outcome $\bar{y}_{qj}$, the mean predicted probability $\bar{\hat{\pi}}_{qj}$ and the number of observations $n_{qj}$.
3. To obtain the pooled estimated risk and observed proportion, we perform a random-effects bivariate meta-analysis using the logit-transformed $\bar{y}_{qj}$ and $\bar{\hat{\pi}}_{qj}$ for each quantile $q$. We use an unstructured covariance matrix[1–3] and estimate both the within-and between-cluster heterogeneity.

Using the fitted model, we estimate the observed proportion and estimated risk of cluster $j$ within quantile $q$. The variance of the random effects (between-cluster variability) as well as the sampling error (within-cluster variability) are captured by the covariance matrices. To fit the model, we utilize the `rma.mv` function of the `metafor` package with cluster as the grouping factor and an unstructured variance-covariance matrix (see https://wviechtb.github.io/metafor/reference/rma.mv.html for a comprehensive overview). Confidence intervals are obtained with profile likelihood, and prediction intervals are calculated as explained in the `metafor` documentation.

**CG-C (interval)**

The algorithm is the same as CG-C (grouped) but in step 1, instead of grouping based on quantiles, we create $Q$ intervals evenly dividing the probability space (0 – 1). Then steps 2 and 3 are identical.

## A2. Two stage meta-analysis (2MA-C)

The two-stage meta-analysis approach combines individual cluster specific calibration models to obtain the calibration in the cluster with the average effect. The process has two stages, first obtaining the individual cluster's calibration and then combining them using random effects meta-analysis as follows:

**Stage 1**

**Fit a flexible calibration model (LOESS or splines, see section 2.3 of the main paper) per cluster and estimate the observed proportion for a grid of values (e.g. 100 values from 0.01 to 0.99).**

We estimate the corresponding observed proportion with the calibration model over a grid ($\boldsymbol{G} = g \in \mathbb{R} \mid 0.01 \leq g \leq 0.99$).

**Stage 2**

**Pool the observed proportion per grid value ($g$) using a random effects model:**

$$logit({}_{s}\hat{\pi}_{gj}) = {}_{\hat{\pi}}\mu_g + \nu_{gj} + \epsilon_{gj}, \quad \epsilon_{gj} \sim N(0, {}_{\hat{\pi}}\sigma^2_{gj}), \quad \nu_{gj} \sim N(0, {}_{\hat{\pi}}\tau^2_g)$$

With $logit({}_s\widehat{\pi}_{gj})$ the logit-transformed predicted probability for point $g$ within cluster $j$, $\nu_{gj}$ the random effect of cluster $j$ and $\epsilon_{gj}$ the error term. The summary estimate is obtained using inverse variance weighting

$$logit({}_s\widehat{\pi}_g) = \frac{\sum_{j=1}^{J} logit({}_s\widehat{\pi}_{gj})\, w_{gj}}{\sum_{j=1}^{J} w_{gj}}$$

where $w_{gj}$ denote the weights calculated as

$$w_{gj} = \frac{1}{{}_{\widehat{\pi}}\tau_g^2 + {}_{\widehat{\pi}}\sigma_{gj}}.$$

${}_{\widehat{\pi}}\tau_g^2$ is the between-cluster variability or heterogeneity estimated with REML (see Veroniki et al. for an overview of methods to estimate ${}_{\widehat{\pi}}\tau_g^2$ [5]) for point $g$ and ${}_{\widehat{\pi}}\sigma_{gj}$ is the within-cluster standard error of point $g$ in cluster $j$. We then use the fitted model per grid value to estimate the observed proportion associated with the grid value and plot the calibration curve.

The confidence interval can be calculated using the Hartung-Knapp-Sidik-Jonkman approach, which is recommended when the number of studies is small [6,7] or with the default method.

Finally we get prediction interval based on the t-distribution as explained in Higgins et al (2009) $logit({}_s\widehat{\pi}_{gj}) \mp t_{J-2}\sqrt{{}_{\widehat{\pi}}\tau_g^2 + SE\big(logit({}_s\widehat{\pi}_{gj})\big)^2}$ [8] (default) where $t_{J-2}$ denotes the t-Student distribution with *J-2* degrees of freedom or any of the supported methods for `meta` package, namely Harktung-Knapp, Kenward Roger [9], bootstrap approach[10] or based on standard normal quantile [11].

## A3. ONE STEP MIXED MODEL (MIX-C)

In this approach, we estimate the observed proportion as

$${}_o\hat{p}_{ij} = \text{logit}^{-1}\left\{\hat{s}\left(\text{logit}\left(\widehat{\pi}(x_{ij})\right)\right) + \widehat{\widehat{s}_j}\left(\text{logit}\left(\widehat{\pi}(x_{ij})\right)\right)\right\}$$

where $\hat{s}$ and $\widehat{\widehat{s}_j}$ denote the estimated smooth effects. We take the variance of both the fixed and random components into account when calculating the variance of the linear predictor and we approximate the standard error of ${}_o\hat{p}_{ij}$ using the delta method. To keep the confidence interval within [0, 1], we construct the interval as

$$min\left(1, max\left(0,\ {}_o\hat{p}_{ij} \mp z_{1-\frac{\alpha}{2}} se\left({}_o\hat{p}_{ij}\right)\right)\right)$$

where $z_{1-\frac{\alpha}{2}}$ denotes the quantile of the standard normal distribution that corresponds to the cumulative probability of $1-\frac{\alpha}{2}$ (i.e. 1.96 for a 95% CI). Prediction intervals are calculated using the *predictInterval* function in R with 10 000 samples (simulation based). This function takes into account the uncertainty at observation level (residual variance), in the fixed coefficients and in the random effects. In this method we first obtain the random and fixed effects, then we generate $n$ samples (default = 10000) based on a multivariate normal distribution of the random and fixed effects, separately. Then we calculate the linear predictor in each sample and predict the upper and lower limits of the prediction interval.

## A4. FIGURES

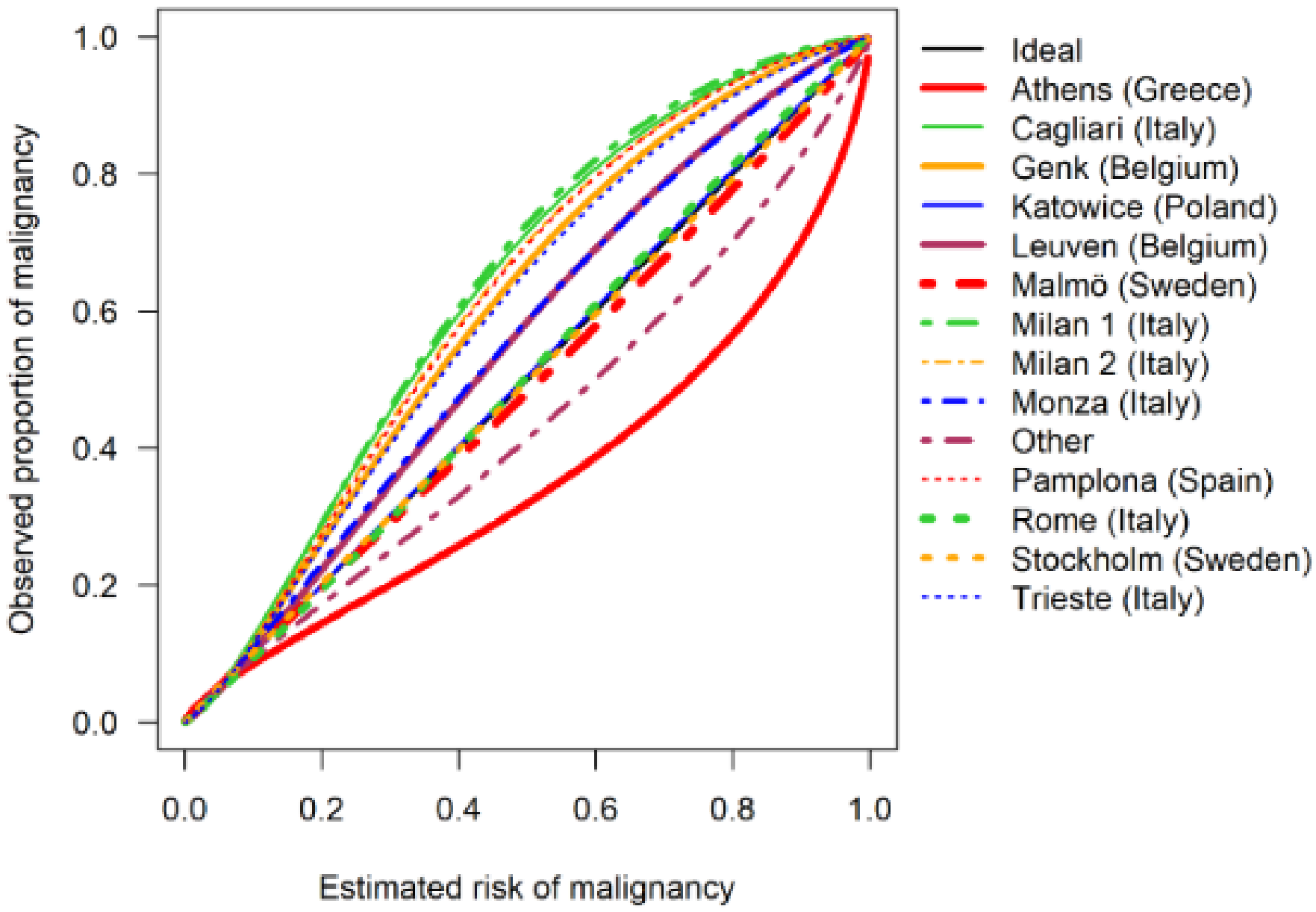

**Figure S1. ADNEX without CA125 center specific logistic calibration curves in IOTA 5 dataset. Reproduced with permission from [12]. Copyright BMJ Publishing Group.**

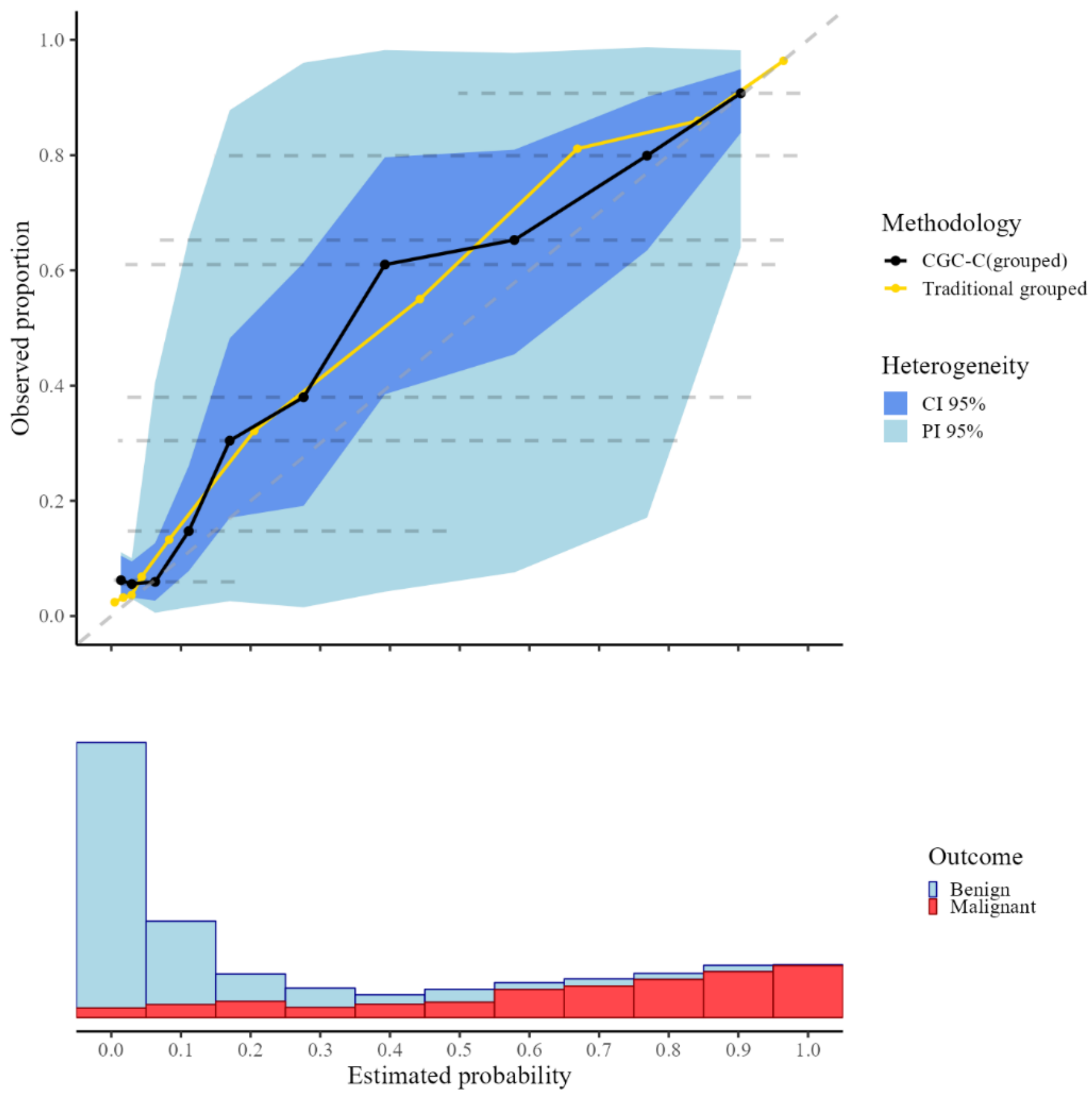

**Figure S2. CG-C (grouped) and traditional grouped calibration plot with 10 quantiles and histogram of estimated risks for the ADNEX model in the motivating example. Dashed horizontal line indicates meta-analysis prediction interval across average estimated probabilities per group.**

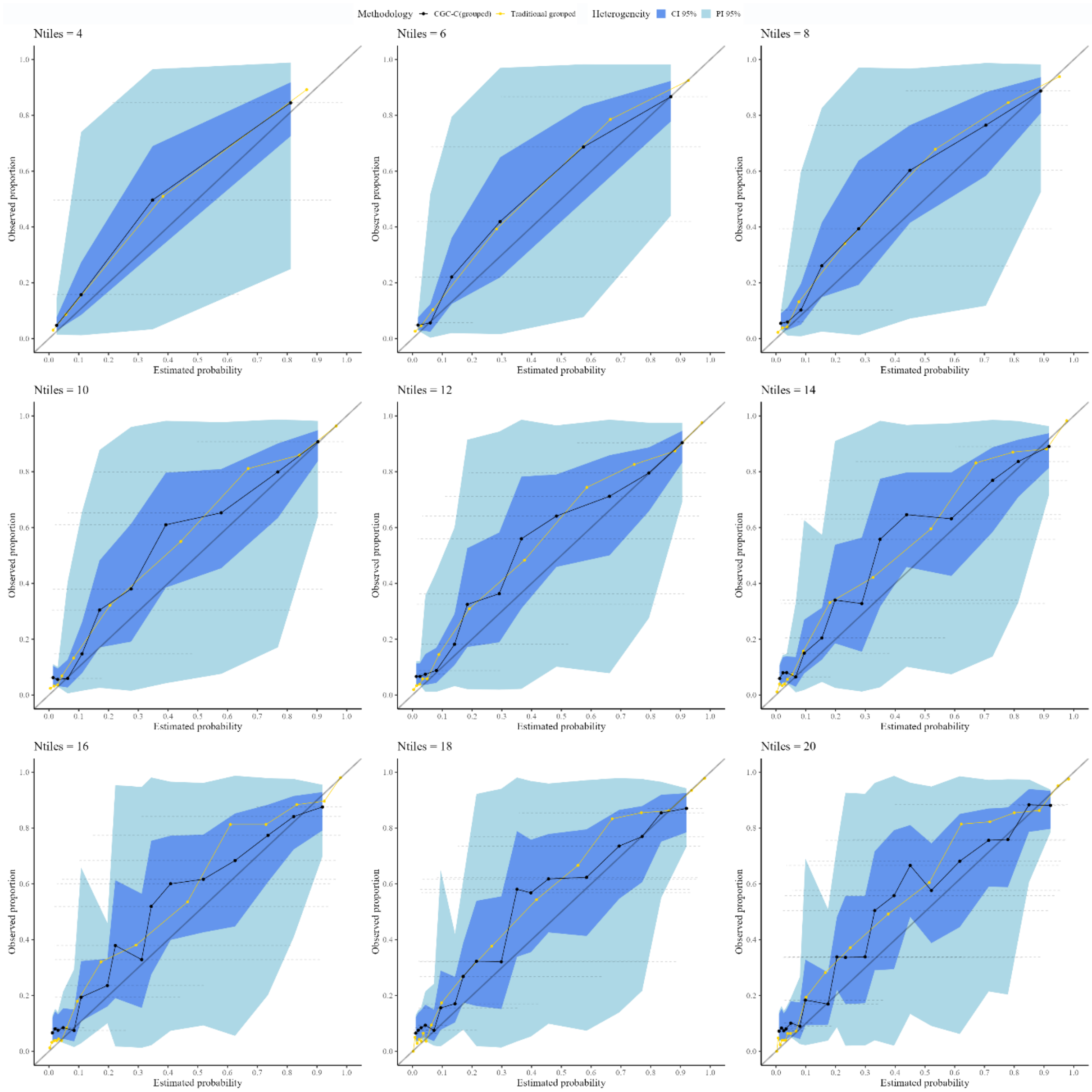

**Figure S3. CG-C (grouped) and traditional grouped calibration plot with varying quantiles from 2 to 20 for the ADNEX model in the motivating example. Dashed horizontal line indicates meta-analysis prediction interval across average estimated probabilities per group.**

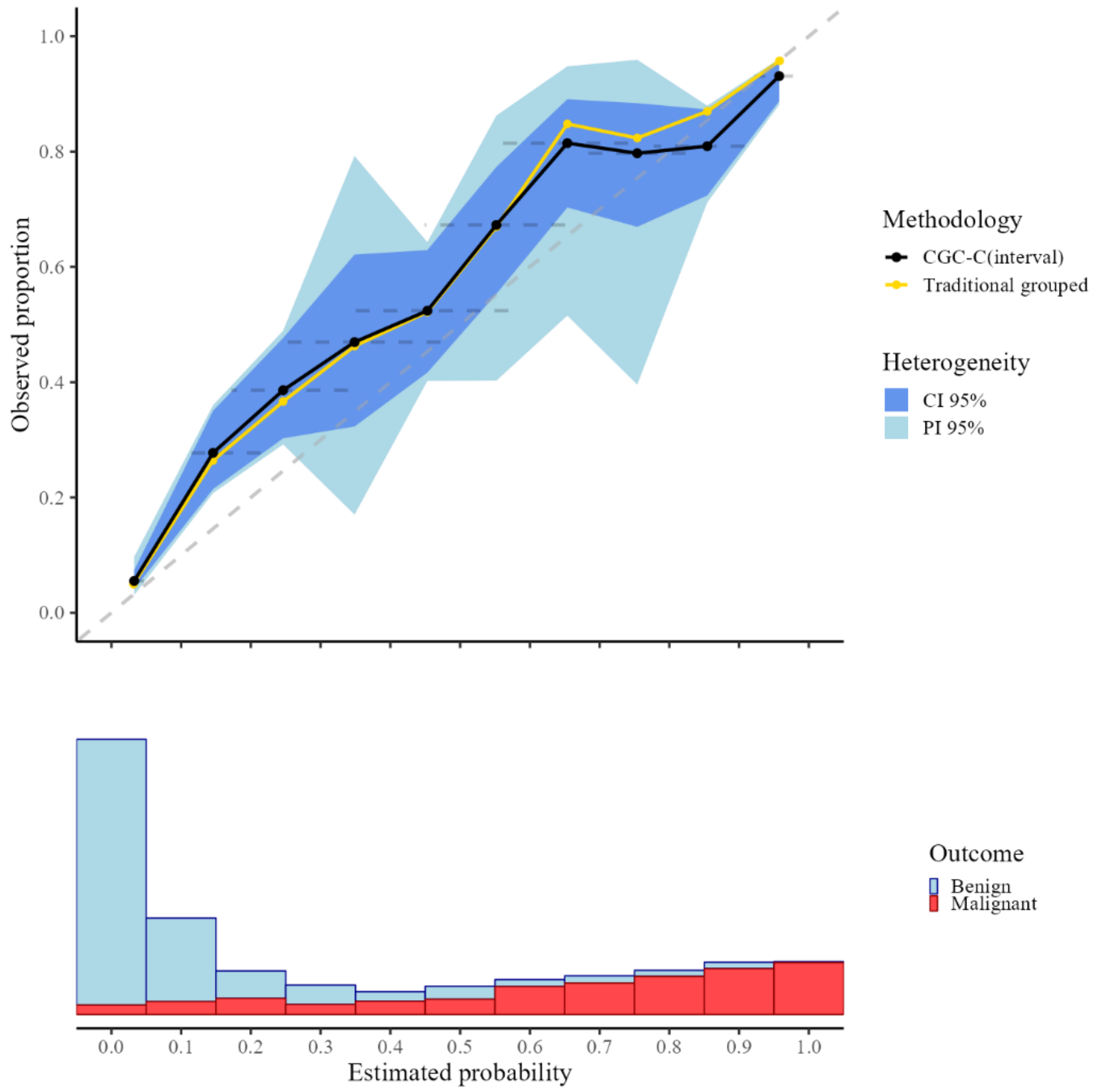

**Figure S4. CG-C (interval) and traditional grouped calibration plot with 10 quantiles and histogram of estimated risks for the ADNEX model in the motivating example. Dashed horizontal line indicates meta-analysis prediction interval across average estimated probabilities per group.**

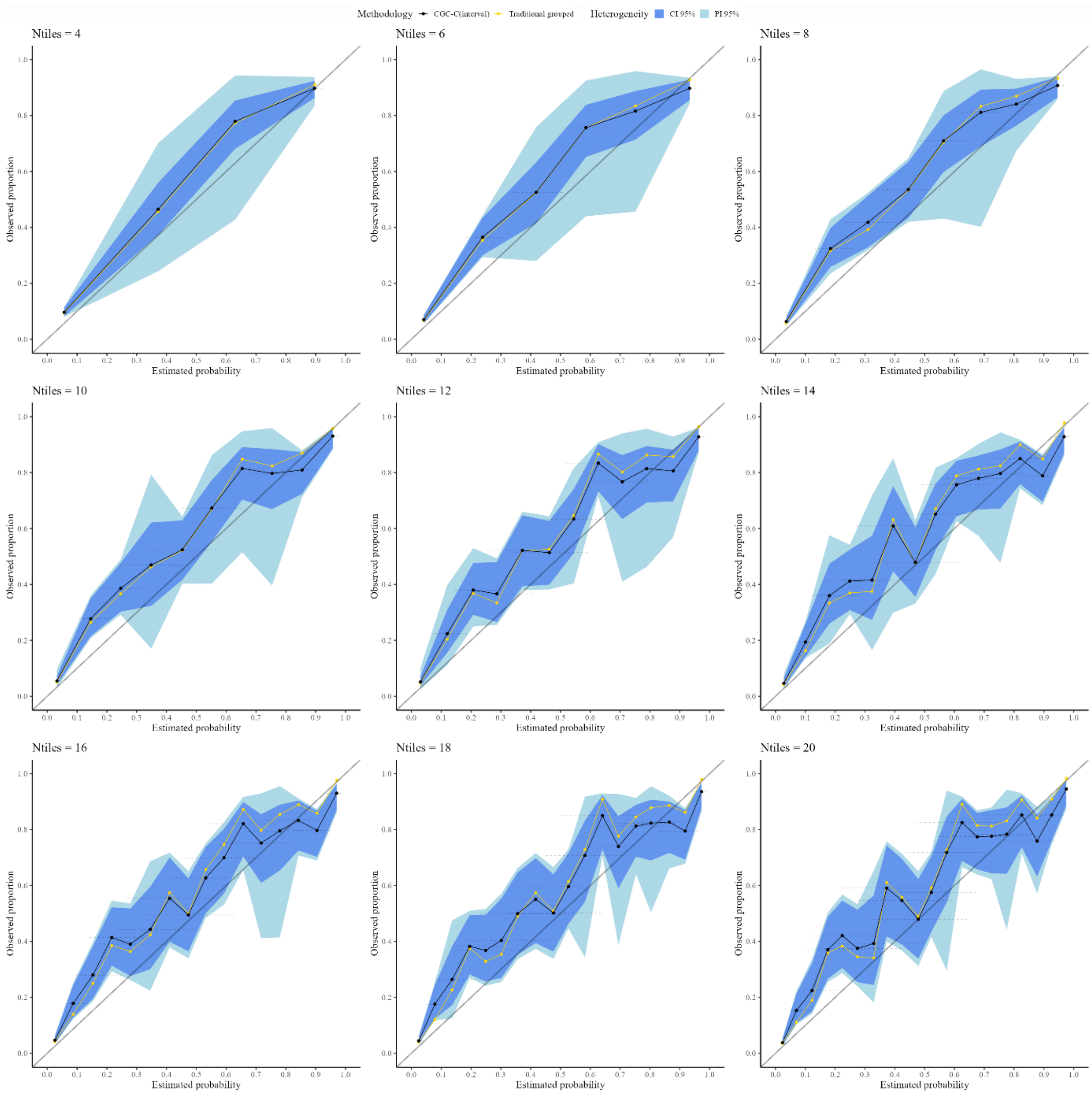

**Figure S5. CG-C (interval) center-specific and traditional grouped calibration plot with varying quantiles from 2 to 20 for the ADNEX model in the motivating example.- Dashed horizontal line indicates meta-analysis prediction interval across average estimated probabilities per group.**

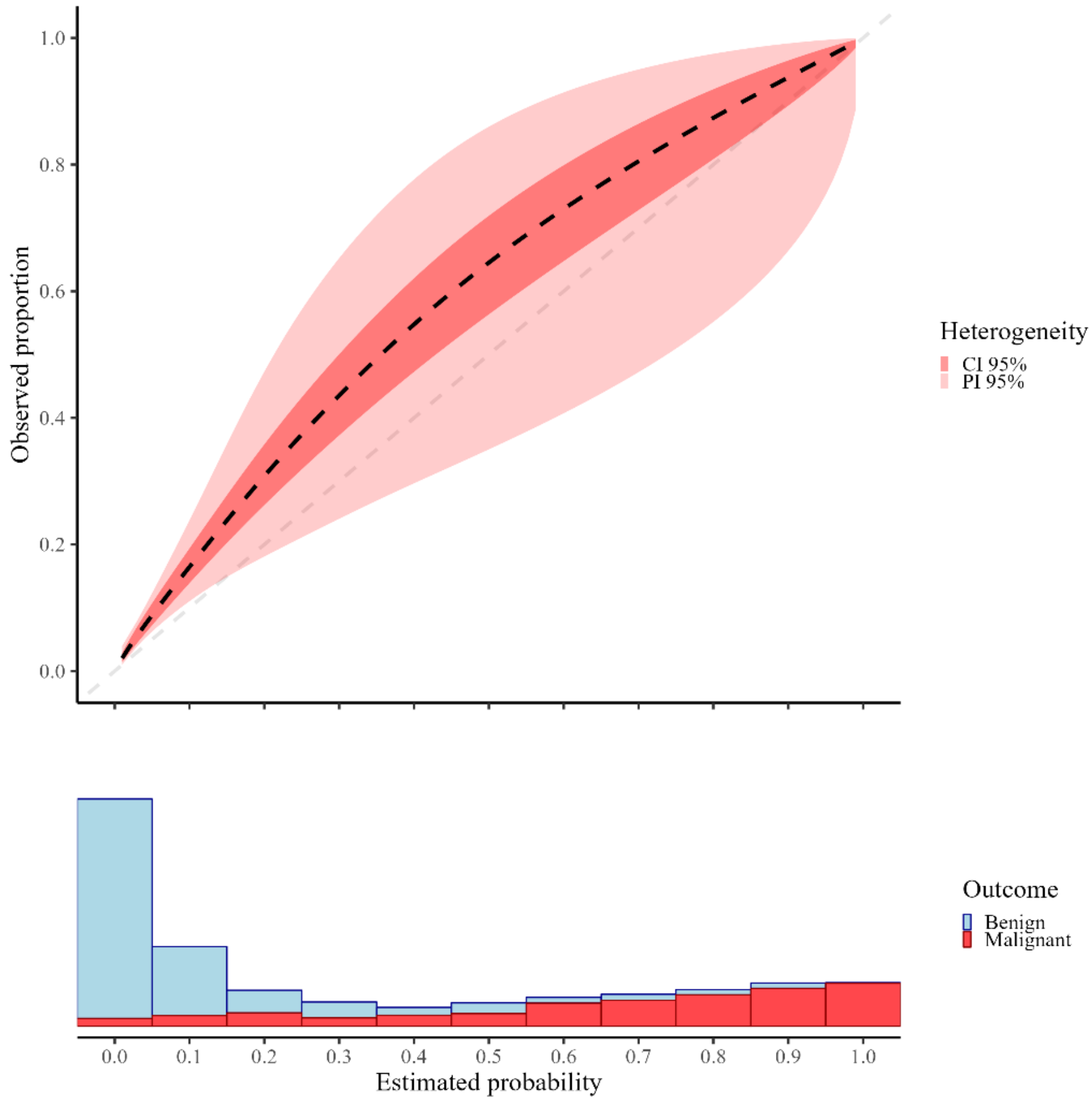

**Figure S6. 2MA-C (splines) calibration plot for the ADNEX model in the motivating example with 100 grid points.**

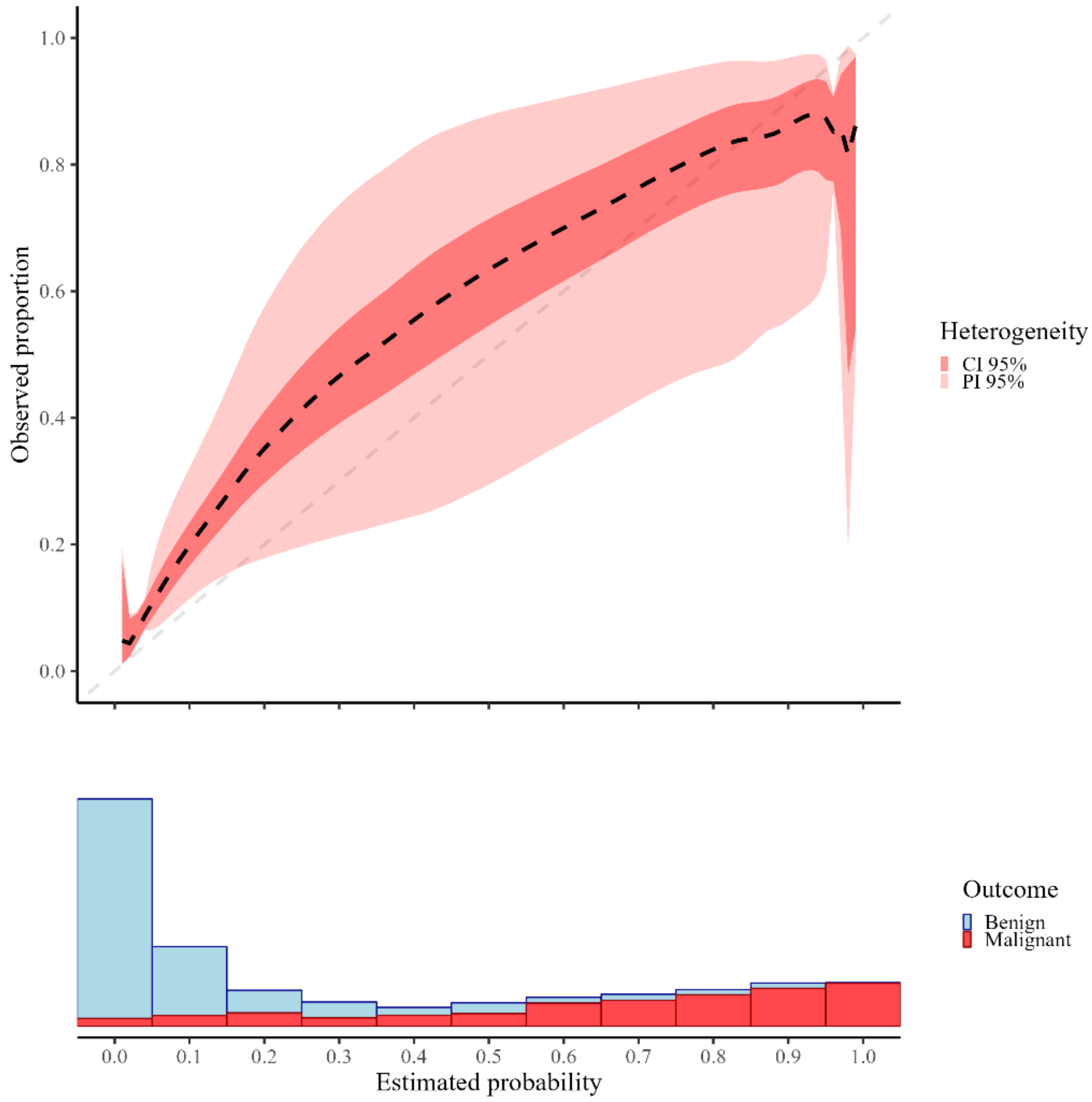

**Figure S7. 2MA-C (loess) calibration plot for the ADNEX model in the motivating example with 100 grid points.**

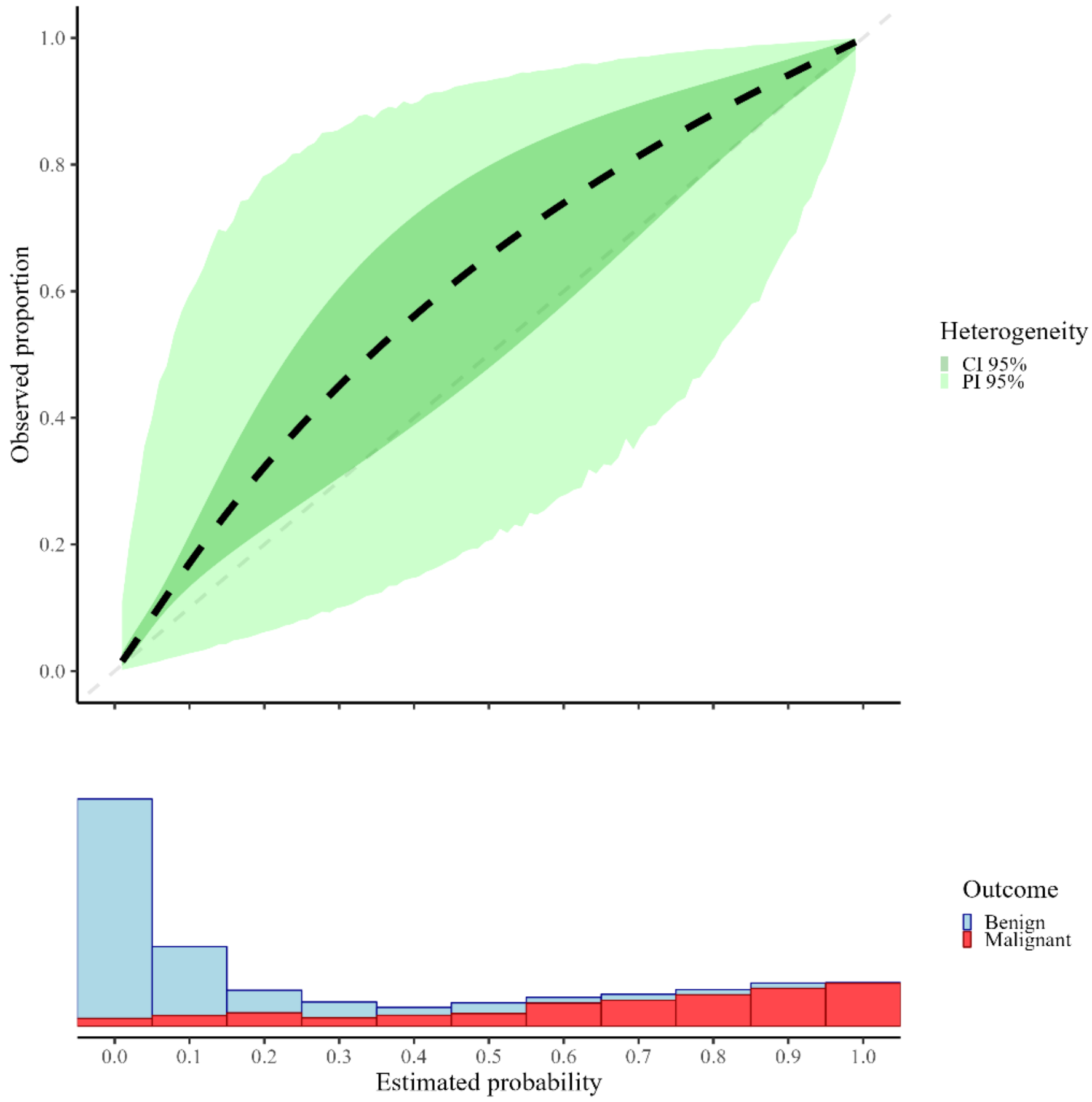

**Figure S8. MIX-C calibration curve based on a model with random intercept and slopes per center and restricted cubic splines for the ADNEX model in the motivating example.**

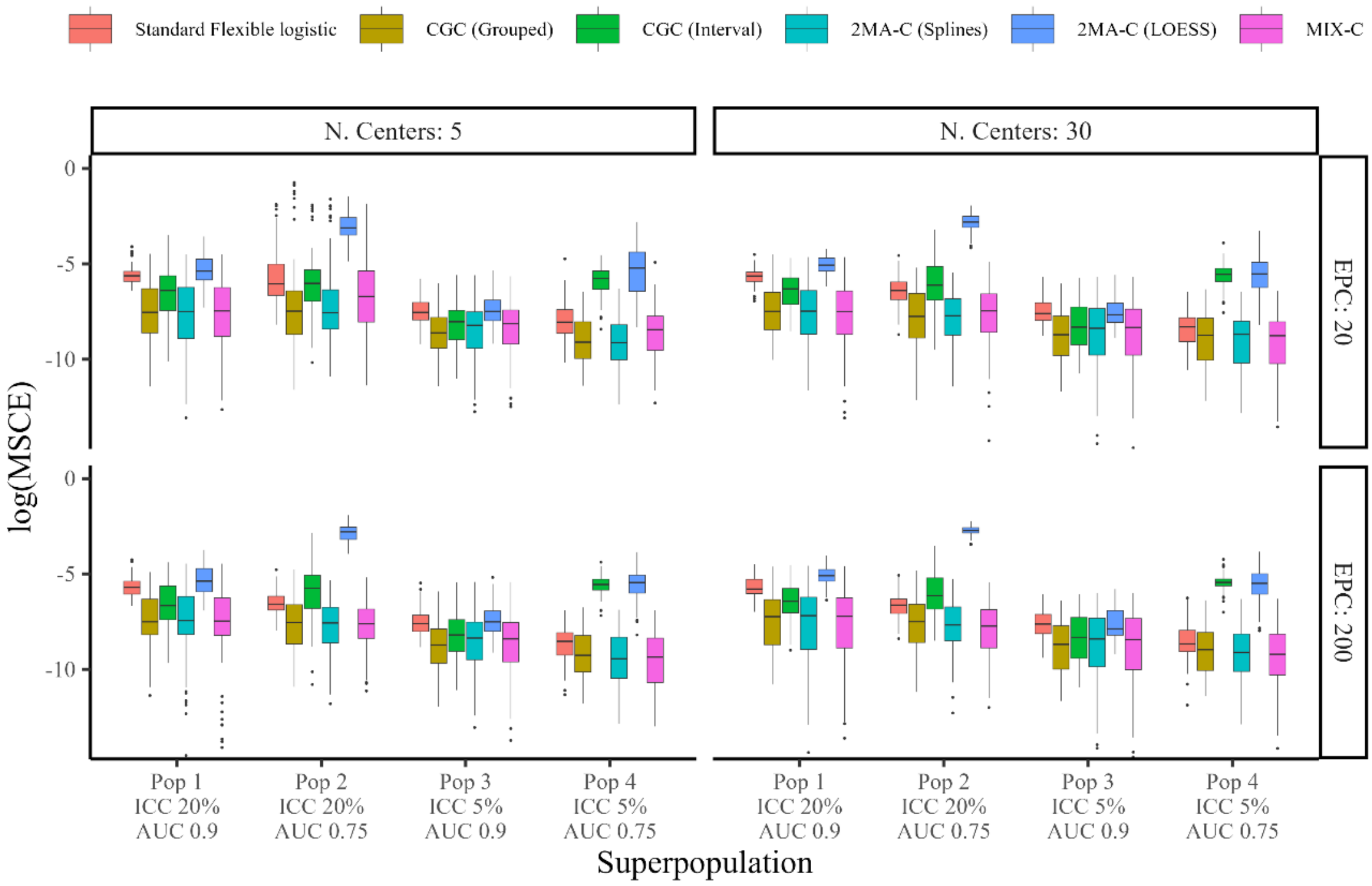

**Figure S9. Boxplots of mean squared calibration error (log) for the prediction model with varying training sample size and fixed validation size of 100.000 patients in 30 centers.**

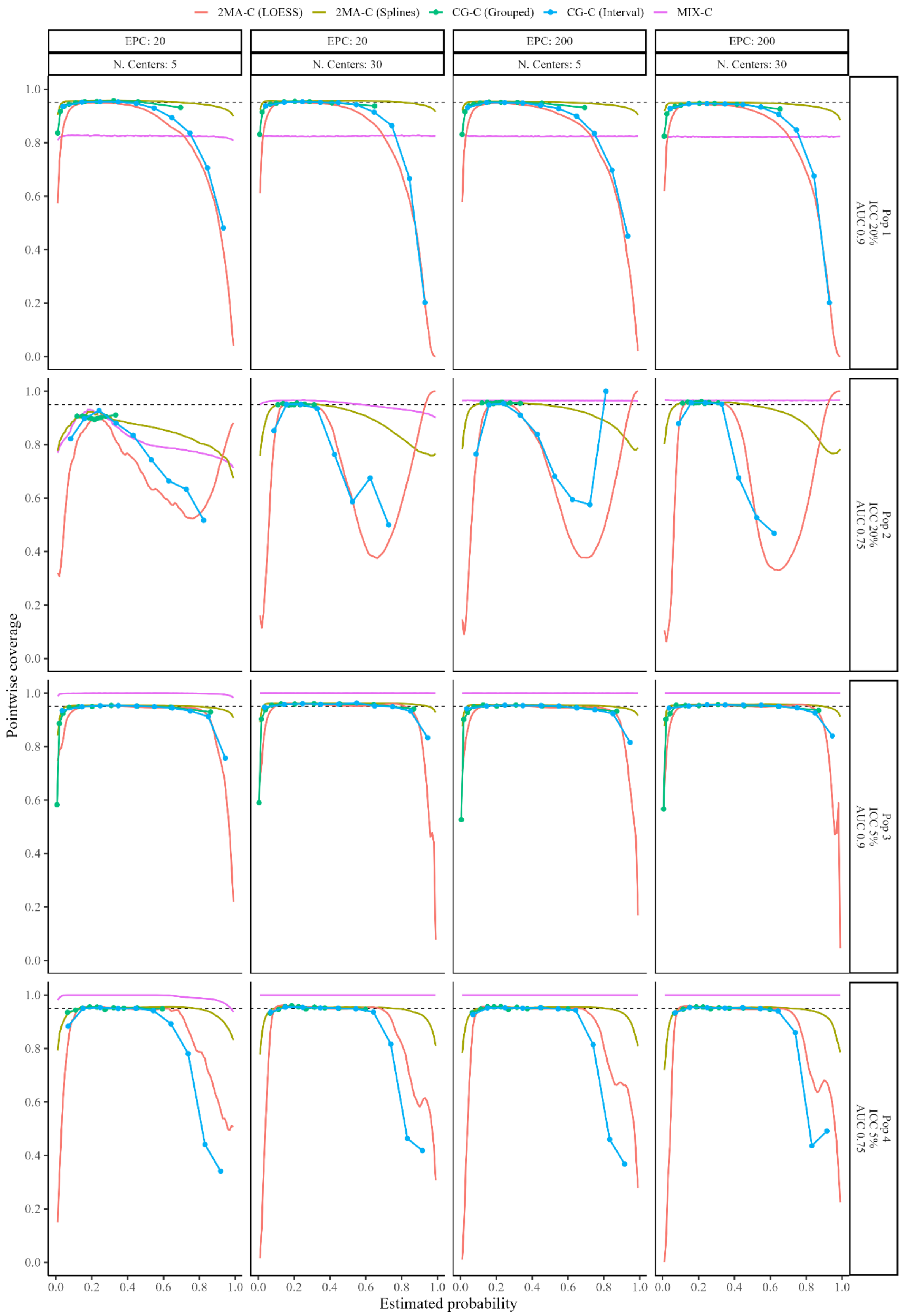

**Figure S10. Pointwise prediction interval coverage across the 16 scenarios with varying training sample size and fixed validation size of 100.000 patients in 30 centers. Black dotted line indicates nominal coverage (95%).**

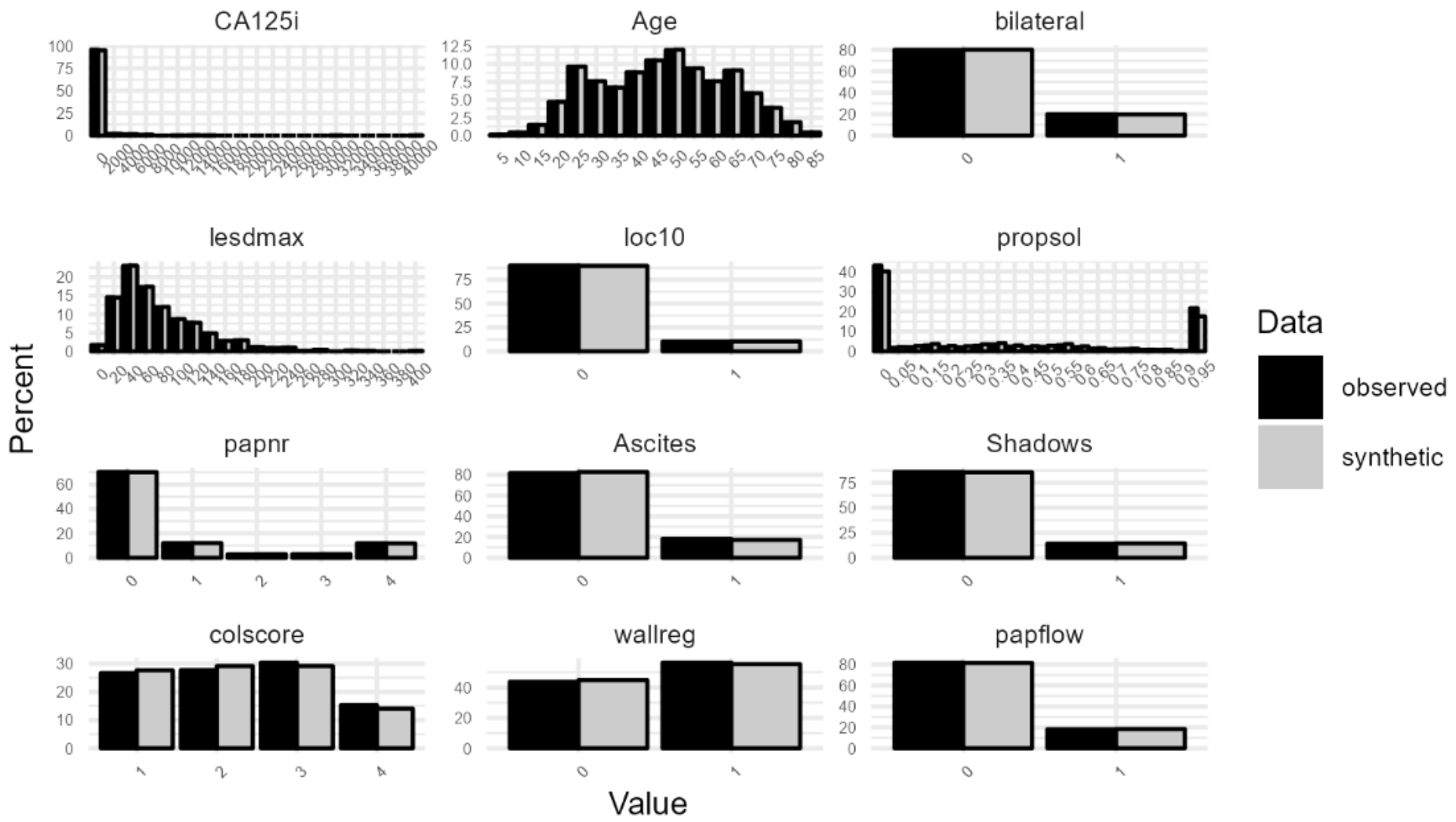

**Figure S11. Quality of synthetic data generation for one of the centers (Leuven).**

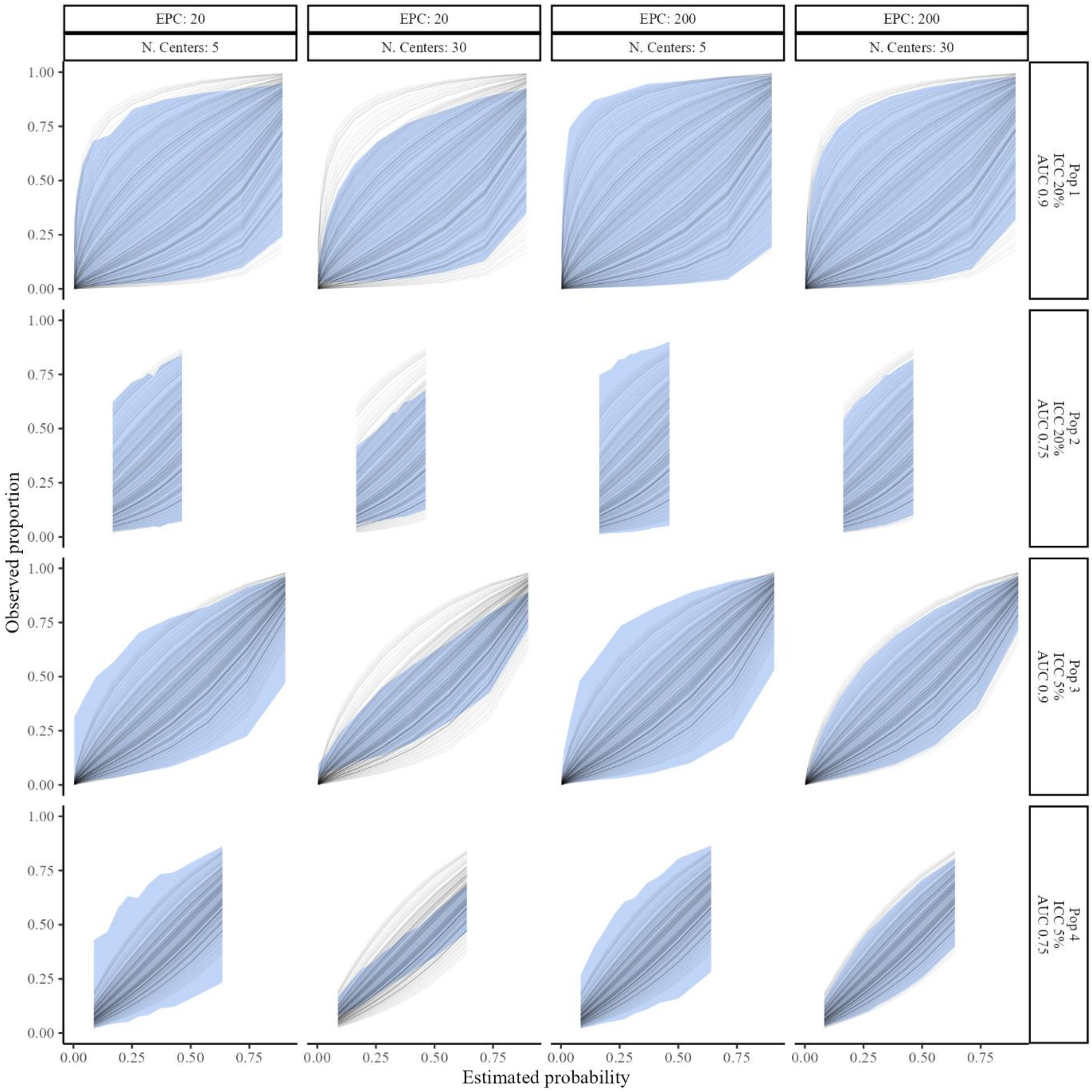

**Figure S12. Average prediction interval across the different scenarios with different validation sample size for the CG-C (grouped) approach. Prediction interval is calculated as the average of the 100 iterations. Black lines represent the true center specific curves in the 200 centers of the superpopulation.**

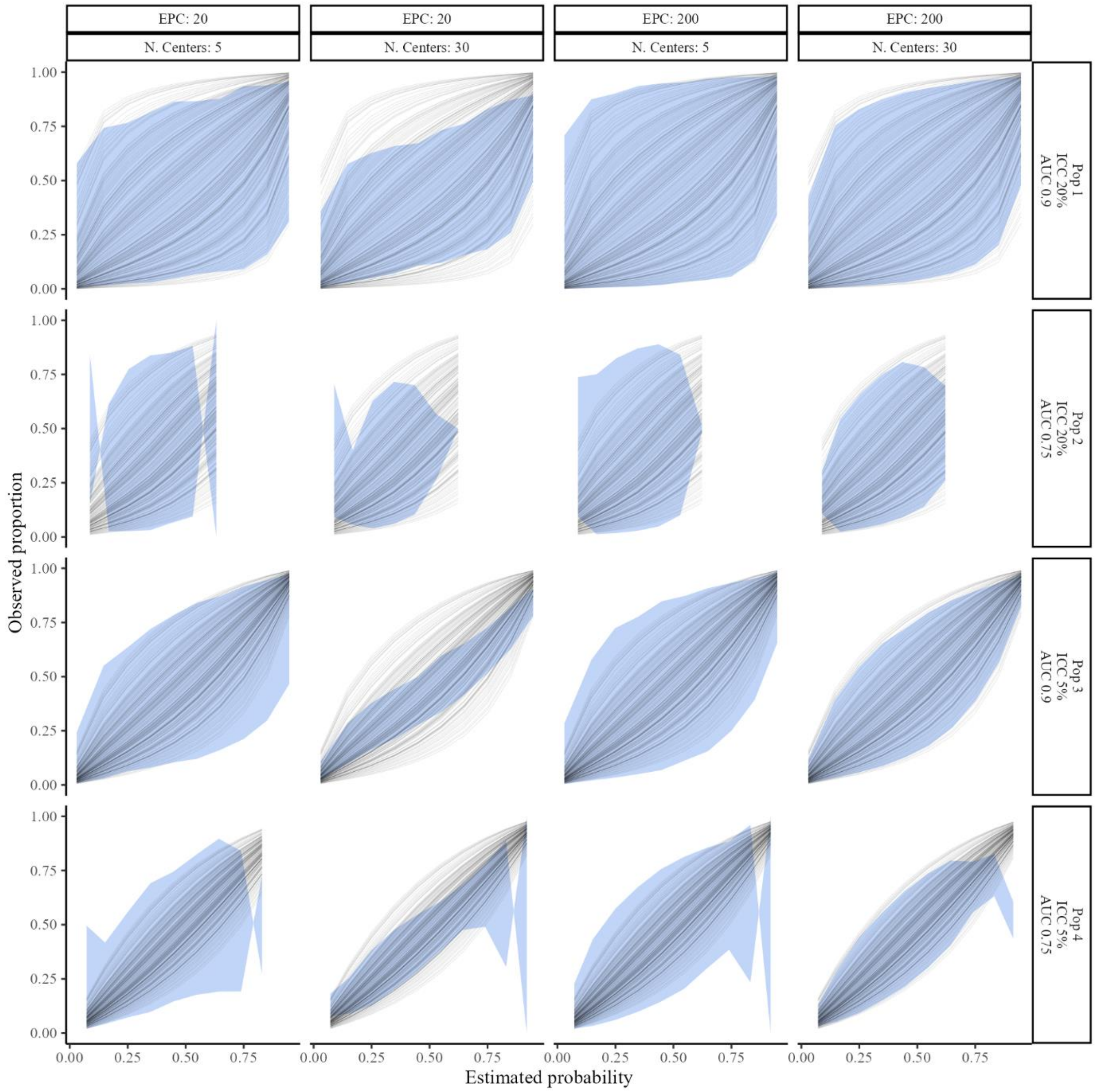

**Figure S13. Average prediction interval across the different scenarios with different validation sample size for the CG-C (interval) approach. Prediction interval is calculated as the average of the 100 iterations. Black lines represent the true center specific curves in the 200 centers of the superpopulation.**

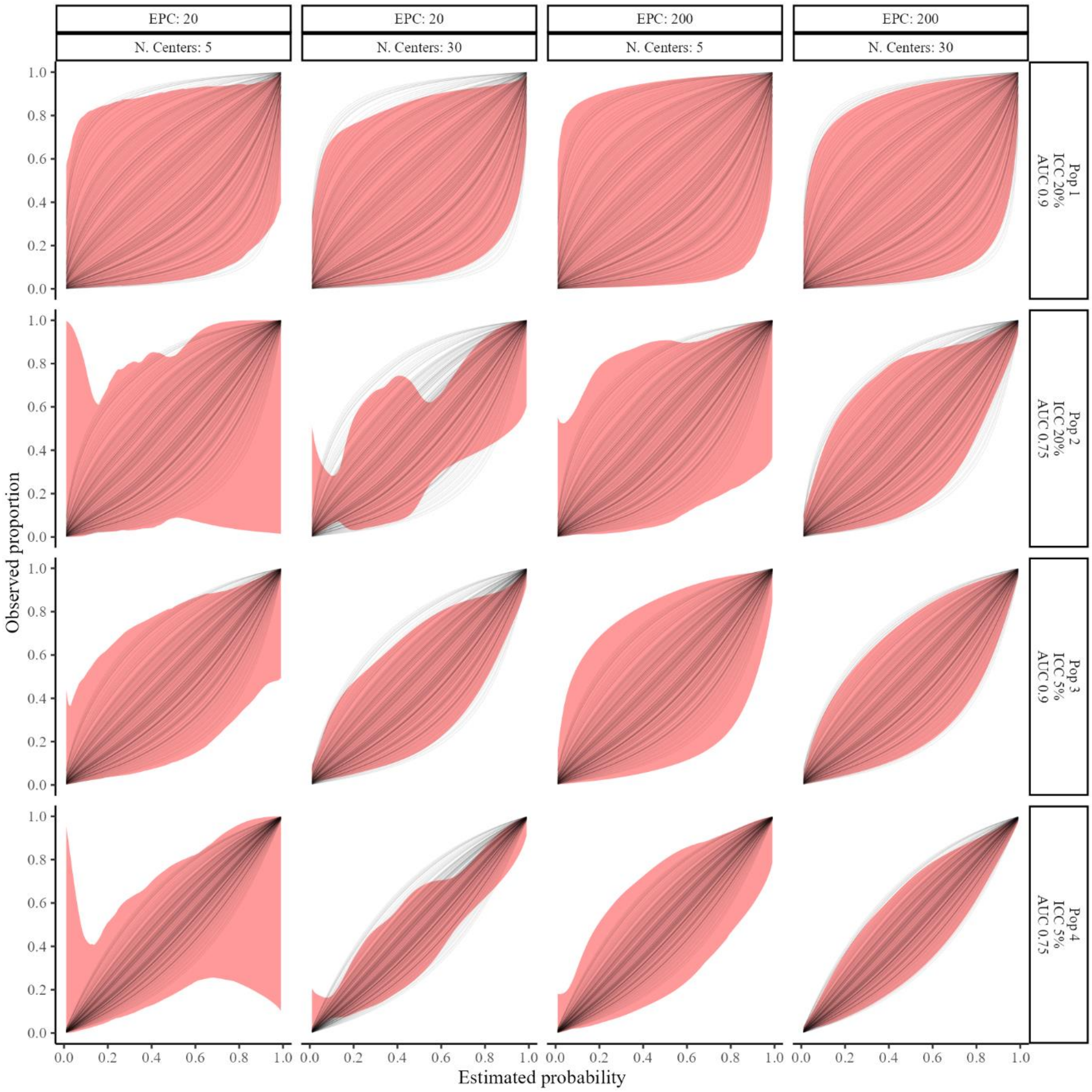

**Figure S14. Average prediction interval across the different scenarios with different validation sample size for the 2MA-C (splines) approach. Prediction interval is calculated as the average of the 100 iterations. Black lines represent the true center specific curves in the 200 centers of the superpopulation.**

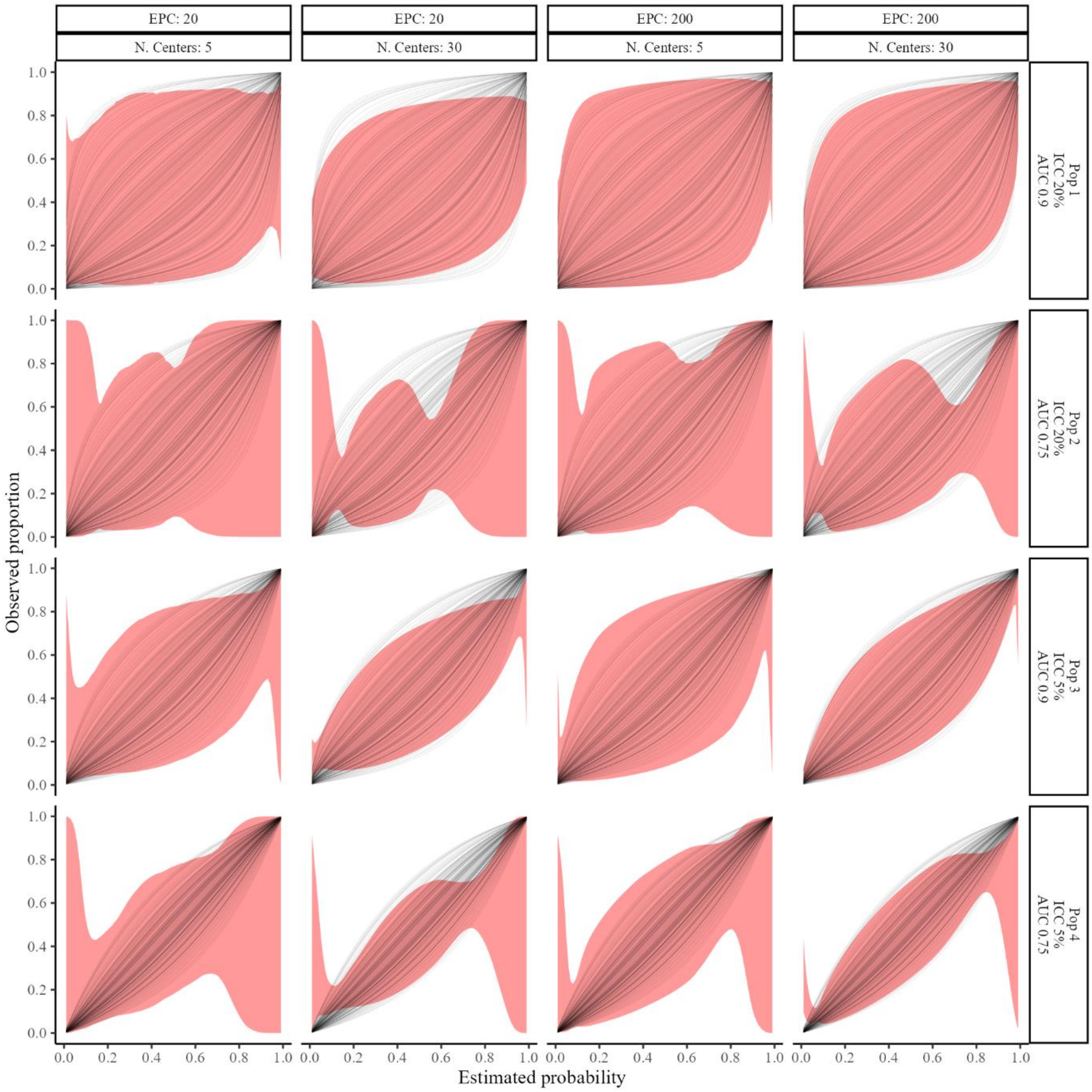

**Figure S15. Average prediction interval across the different scenarios with different validation sample size for the 2MA-C (LOESS) approach. Prediction interval is calculated as the average of the 100 iterations. Black lines represent the true center specific curves in the 200 centers of the superpopulation.**

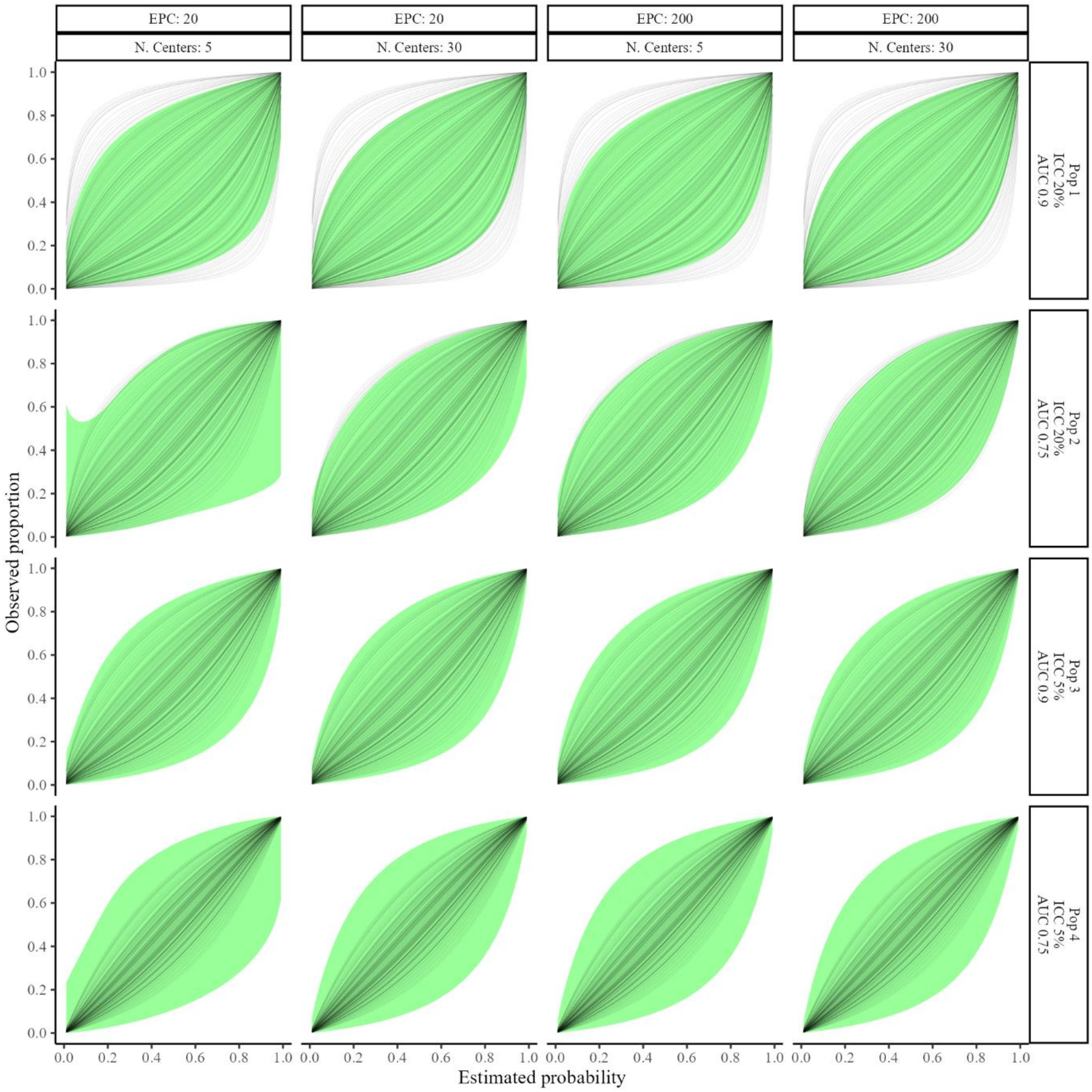

**Figure S16. Average prediction interval across the different scenarios with different validation sample size for the MIX-C approach. Prediction interval is calculated as the average of the 100 iterations. Black lines represent the true center specific curves in the 200 centers of the superpopulation.**

## A5. TABLES

| Superpopulation | AUC | ICC | Event rate (range) | Formula |
|---|---|---|---|---|
| P1 | 0.9 | 0.2 | 0.3 (0.04-0.74) | $\pi(\boldsymbol{x}_{ij}) = -1.6054 - 2.09062x_{ij} + u_j$ <br> $u_j \sim N(0,1.559)$ |
| P2 | 0.75 | 0.2 | 0.3 (0.05-0.75) | $\pi(\boldsymbol{x}_{ij}) = -1.0122 + 0.4199x_{ij} + u_j$ <br> $u_j \sim N(0,1.0024)$ |
| P3 | 0.9 | 0.05 | 0.3 (0.14-0.51) | $\pi(\boldsymbol{x}_{ij}) = -1.5943 + 2.3875x_{ij} + u_j$ <br> $u_j \sim N(0,0.7827)$ |
| P4 | 0.75 | 0.05 | 0.3 (0.14-0.52) | $\pi(\boldsymbol{x}_{ij}) = -1.0244 - 0.9273x_{ij} + u_j$ <br> $u_j \sim N(0,0.5183)$ |

**Table S1. Superpopulation characteristics**

## REFERENCES

1. van Houwelingen HC, Arends LR, Stijnen T. Advanced methods in meta-analysis: multivariate approach and meta-regression. *Stat Med*. 2002;21(4):589-624. doi:10.1002/sim.1040

2. Berkey CS, Hoaglin DC, Mosteller F, Colditz GA. A random-effects regression model for meta-analysis. *Stat Med*. 1995;14(4):395-411. doi:10.1002/sim.4780140406

3. Reitsma JB, Glas AS, Rutjes AWS, Scholten RJPM, Bossuyt PM, Zwinderman AH. Bivariate analysis of sensitivity and specificity produces informative summary measures in diagnostic reviews. *J Clin Epidemiol*. 2005;58(10):982-990. doi:10.1016/j.jclinepi.2005.02.022

4. Riley RD, Higgins JPT, Deeks JJ. Interpretation of random effects meta-analyses. *BMJ*. 2011;342:d549. doi:10.1136/bmj.d549

5. Veroniki AA, Jackson D, Viechtbauer W, et al. Methods to estimate the between‐study variance and its uncertainty in meta‐analysis. *Res Synth Methods*. 2016;7(1):55-79. doi:10.1002/jrsm.1164

6. Schwarzer G, Carpenter JR, Rücker G. *Meta-Analysis with R*. Cham: Springer International Publishing; 2015. doi:10.1007/978-3-319-21416-0

7. Snell KI, Ensor J, Debray TP, Moons KG, Riley RD. Meta-analysis of prediction model performance across multiple studies: Which scale helps ensure between-study normality for the C-statistic and calibration measures? *Stat Methods Med Res*. 2018;27(11):3505-3522. doi:10.1177/0962280217705678

8. Higgins JPT, Thompson SG, Spiegelhalter DJ. A re-evaluation of random-effects meta-analysis. *J R Stat Soc Ser A Stat Soc*. 2009;172(1):137-159. doi:10.1111/j.1467-985X.2008.00552.x

9. Partlett C, Riley RD. Random effects meta-analysis: Coverage performance of 95% confidence and prediction intervals following REML estimation. *Stat Med*. 2017;36(2):301-317. doi:10.1002/sim.7140

10. Nagashima K, Noma H, Furukawa TA. Prediction intervals for random-effects meta-analysis: A confidence distribution approach. *Stat Methods Med Res*. 2019;28(6):1689-1702. doi:10.1177/0962280218773520

11. Skipka G. The inclusion of the estimated inter-study variation into forest plots for random effects meta-analyses – a suggestion for a graphical representation. https://abstracts.cochrane.org/2006-dublin/inclusion-estimated-inter-study-variation-forest-plots-random-effects-meta-analyses. Published 2006.

12. Van Calster B, Valentin L, Froyman W, et al. Validation of models to diagnose ovarian cancer in patients managed surgically or conservatively: multicentre cohort study. *BMJ*. 2020;370:m2614. doi:10.1136/bmj.m2614